\documentclass[a4paper,fleqn,usenatbib]{mnras}

\usepackage{newtxtext,newtxmath}
\usepackage[T1]{fontenc}
\usepackage{ae,aecompl}

\usepackage{graphicx}	% Including figure files
\usepackage{amsmath}	% Advanced maths commands
\newcommand{\angstrom}{\textup{\AA}}

\defcitealias{Mathis:1977aa}{MRN}
\defcitealias{Hirashita:2017aa}{HH17}
\defcitealias{Hirashita:2020aa}{HM20}
\defcitealias{Glover:2011aa}{GM11}

\title[Dust and molecules in galaxies]{Effects of dust grain size distribution on the abundances
of CO and H$_2$ in galaxy evolution}

\author[H. Hirashita]{
Hiroyuki Hirashita$^{1,2}$\thanks{E-mail: hirashita@asiaa.sinica.edu.tw}
\\
$^{1}$Institute of Astronomy and Astrophysics, Academia Sinica,
Astronomy-Mathematics Building, No.\ 1, Section 4,
Roosevelt Road, Taipei 10617, Taiwan\\
$^{2}$Theoretical Astrophysics, Department of Earth and Space Science, Osaka University,
1-1 Machikaneyama, Toyonaka, Osaka 560-0043, Japan
}

\date{Accepted XXX. Received YYY; in original form ZZZ}

\pubyear{2023}

\begin{document}
\label{firstpage}
\pagerange{\pageref{firstpage}--\pageref{lastpage}}
\maketitle

\begin{abstract}
We model the effect of grain size distribution in a galaxy on the evolution of CO and
H$_2$ abundances.
The formation and dissociation of CO and H$_2$ in typical dense clouds
are modelled in a manner consistent with
the grain size distribution.
The evolution of grain size distribution is calculated based on our previous model,
which treats the galaxy as a one-zone object but includes various dust
processing mechanisms in the interstellar medium (ISM).
We find that typical dense clouds become fully molecular (H$_2$) when the dust surface area
increases by shattering while an increase of dust abundance
by dust growth in the ISM is necessary for a significant rise of the CO abundance.
Accordingly, the metallicity dependence of the CO-to-H$_2$ conversion factor,
$X_\mathrm{CO}$, is predominantly driven by dust growth.
We also examine the effect of grain size distribution in the galaxy by changing the dense gas fraction,
which controls the balance between coagulation and shattering, clarifying that the
difference in the grain size distribution significantly affects $X_\mathrm{CO}$ even if the
dust-to-gas ratio is the same. The star formation time-scale, which controls the speed
of metal enrichment also affects the metallicity at which the CO abundance rapidly increases
(or $X_\mathrm{CO}$ drops). We also propose dust-based formulae for $X_\mathrm{CO}$,
which need further tests for establishing their usefulness.
\end{abstract}

\begin{keywords}
molecular processes -- dust, extinction -- galaxies: evolution -- galaxies: ISM --
radio lines: galaxies
\end{keywords}

%%%%%%%%%%%%%%%%%%%%%%%%%%%%%%%%%%%%%%%%%%%%%%%%%%

%%%%%%%%%%%%%%%%% BODY OF PAPER %%%%%%%%%%%%%%%%%%

\section{Introduction}

Star-forming clouds are usually rich in molecules, and hydrogen nuclei are in the form of
H$_2$ in those clouds, which are thus referred to as molecular clouds.
Since H$_2$ does not efficiently emit in low-temperature environments,
carbon monoxide (CO) is often used as a tracer of molecular clouds.
In the present Universe, H$_2$ forms predominantly on dust surfaces
\citep{Gould:1963aa,Cazaux:2004aa} while
CO forms through gas-phase chemical reactions. For the formation of both molecular species,
a condition shielded from dissociating ultraviolet (UV) radiation is favourable, requiring a
high column density.
H$_2$ molecules shield UV radiation by their own absorption \citep{Draine:1996aa},
which is referred to as self-shielding.
Both molecular species are also
influenced by dust, which also shields UV radiation efficiently (this mechanism
is referred to as dust shielding).

To infer the H$_2$ mass from the observed CO line strength, we usually assume
a CO-to-H$_2$ conversion factor, $X_\mathrm{CO}\equiv N_\mathrm{H_2}/W_\mathrm{CO}$,
where $N_\mathrm{H_2}$ is the H$_2$ column density,
and $W_\mathrm{CO}$ is the CO
$J=1\to 0$ emission line intensity integrated for the frequency (often expressed
by the Doppler shift velocity in units of km~s$^{-1}$). The conversion factor can also be
expressed as $\alpha_\mathrm{CO}\equiv\Sigma_\mathrm{mol}/W_\mathrm{CO}$
based on
the surface mass density of the molecular
gas, $\Sigma_\mathrm{mol}=1.36m_\mathrm{H}(2N_\mathrm{H_2})$,
where the factor 1.36 accounts for the contribution of helium.

For the purpose of obtaining the CO-to-H$_2$ conversion factor,
the H$_2$ surface density or the mass of a molecular cloud needs to be
estimated through the virial theorem (i.e.\ a dynamical mass estimate)
or a conversion from the dust far-infrared intensity to the total gas column density
\citep[see][for a review]{Bolatto:2013aa}. The obtained H$_2$ content, compared with the
CO line intensity, leads to an estimate of $X_\mathrm{CO}$, which has been derived for various
nearby galaxies. Many studies have found that $X_\mathrm{CO}$ strongly
depends on the metallicity
\citep{Wilson:1995aa,Arimoto:1996aa,Israel:1997aa,Bolatto:2008aa,Leroy:2011aa,Hunt:2015aa}.
Similar metallicity dependence of $X_\mathrm{CO}$ is also observed for galaxies
at $z\sim 1$--3, where $z$ is the redshift \citep{Genzel:2012aa}.
It is also theoretically expected that $X_\mathrm{CO}$ depends on the gas density and
temperature \citep{Feldmann:2012aa,Narayanan:2012aa}.

The above metallicity dependence can be interpreted as a result of more dust shielding
of CO-dissociating radiation at higher metallicity, since the abundances of dust and metals
are strongly related to each other \citep{Issa:1990aa,Schmidt:1993aa,Lisenfeld:1998aa,Dwek:1998aa}.
As mentioned above, dust influences both CO and
H$_2$ abundances via shielding of UV dissociating
photons and formation of H$_2$ on grain surfaces.
For these processes,
the cross-section for UV radiation and the grain surface area
are important, and are
determined not only by the dust abundance, but also by
the grain size distribution \citep{Yamasawa:2011aa}.
Therefore, to clarify the above metallicity dependence of $X_\mathrm{CO}$,
we need to understand how the grain size distribution as well as the dust abundance
evolves as a function of metallicity.

There are some theoretical models for the evolution of grain size distribution
(as well as that of total dust abundance) in galaxies.
\citet{Asano:2013aa} modelled the evolution of
grain size distribution in a manner consistent with the metal enrichment of a galaxy.
This model is later modified by \citet{Nozawa:2015aa}, who better explained the Milky Way (MW)
extinction curve by including stronger dust growth that happens in dense molecular clouds.
In their models, the grain size distribution evolves in the following way:
In the early epoch, the galaxy is enriched with dust by stellar dust production
(dust condensation in stellar ejecta), and the grain size distribution is dominated by
large submicron-sized  grains. As the dust abundance increases, grain--grain collisions
in the diffuse interstellar medium (ISM) become frequent enough for the grains to be shattered.
The formed small grains efficiently accrete the surrounding gas-phase metals in the
dense ISM because of their large surface area. This process, referred to as
accretion, drastically increases the abundance of small grains, and the total dust mass.
Afterwards, the small grains are coagulated into large grains
in the dense ISM. As a consequence, in solar-metallicity environments, the grain size
distribution tends to converge to a power-law shape similar to the one derived by
\citet[][hereafter MRN]{Mathis:1977aa} because of the balance between shattering
and coagulation.

\citet[][hereafter HH17]{Hirashita:2017aa} developed a theoretical model to
investigate how the evolution of grain size distribution influences
the abundances of H$_2$ and CO molecules. To save the computational load,
they adopted the two-size approximation formulated by \citet{Hirashita:2015ab}.
In this approach, the full grain radius range is approximated with
two bins separated
around a radius of 0.03~$\micron$. They evaluated the grain surface reaction rate for H$_2$
and the shielding of UV dissociating radiation analytically,
taking into account the information on the grain size distribution in the form of
small-to-large grain abundance ratio.
\citetalias{Hirashita:2017aa} found that, among the processes involved in dust evolution,
dust growth by accretion plays
the most important role in increasing the CO abundance and decreasing $X_\mathrm{CO}$
while the increase of H$_2$ occurs even before dust growth takes place significantly
\citep[see also][]{Hu:2023aa}.
Moreover, they also clarified that the difference in small-to-large grain abundance ratio
has a large impact on the shielding of UV dissociating radiation.
As a consequence, $X_\mathrm{CO}$ could be different by an order of magnitude
depending on the grain size distribution even if the dust abundance is the same.
This result underlines the importance of the
grain size distribution in estimating the molecular gas content
from CO emission.

The \citetalias{Hirashita:2017aa} model was also useful to calculate the spatial distributions of
H$_2$ and CO in a galactic disc \citep{Chen:2018aa}
by post-processing an isolated disc galaxy simulation
by \citet{Aoyama:2017aa} and \citet{Hou:2017aa}.
\citet{Chen:2018aa} showed that
the relation between star formation rate (SFR) and H$_2$ or CO surface density
is strongly affected by the grain size distribution.
Therefore, appropriately modelling the H$_2$ and CO abundances in a manner
consistent with the grain size distribution is important in the `star formation law' --
the relation between the surface densities of H$_2$ or CO mass and SFR.

Although the two-size approximation adopted by \citetalias{Hirashita:2017aa} is useful
for analytically calculating the H$_2$ and CO abundances,
it has only two degrees of freedom (large and small grains)
in predicting the extinction curve and the grain surface area, which are important for
shielding of dissociating radiation and H$_2$ formation on grain surfaces, respectively.
Now using the full grain size distribution instead of the two size approximation is a natural extension of
our previous studies. In fact, the evolution of grain size distribution is complex
and strongly time-dependent especially at subsolar metallicity, where dust growth by accretion
drastically increases the abundance of small grains \citep{Asano:2013aa}.
Since the change of $X_\mathrm{CO}$ is also large at subsolar metallicity, it is important to
catch the evolution of grain size distribution correctly.
Therefore, in this paper, we aim at consistently modelling the H$_2$ and CO abundances
with the evolution of grain size distribution. This step is useful in the following two points:
(i) We are able to check if the previous results with the two-size approximation reasonably predicted
the evolution of the molecular abundances and $X_\mathrm{CO}$.
(ii) The framework developed in this paper can be used to predict the H$_2$ and CO abundances
in hydrodynamic simulations that include the grain size distribution. Recently,
some hydrodynamic simulations have succeeded in implementing the evolution of grain
size distribution \citep{McKinnon:2018aa,Aoyama:2020aa,Li:2021ab,Romano:2022ab}.
Some galaxy-scale simulations treated
H$_2$ formation in a manner consistent with the evolution of dust abundance and showed that
dust evolution plays an important role in H \textsc{i}--H$_2$ transition
\citep{Bekki:2013aa,Osman:2020aa}. However, these studies did not include the
evolution of grain size distribution.
\citet{Romano:2022aa} have also calculated the H$_2$ abundance
in their hydrodynamic simulation of an isolated galaxy, which incorporated the evolution of
grain size distribution, but they have not yet calculated the CO abundance.
This simulation, if
combined with our models to be developed in this paper, would enable us to obtain
spatially resolved H$_2$ and CO maps in galaxies.

The goal of this paper is to model the H$_2$ and CO abundances in dense clouds
in a manner consistent with the evolution of grain size distribution in the galaxy hosting these clouds.
This predicts not only the evolution of the H$_2$ and CO abundances in dense clouds
but also that of $X_\mathrm{CO}$. Given that dust enrichment is strongly related
to metallicity increase, we predict the metallicity dependence of $X_\mathrm{CO}$,
which is compared with observations. We also focus on some parameters that control
the grain size distribution; this procedure serves to clarify the effect of grain size distribution
on the molecular abundances.

This paper is organized as follows.
In Section \ref{sec:model}, we review the dust evolution model,
and explain the calculation method for the abundances
of H$_2$ and CO, and the CO-to-H$_2$ conversion factor.
In Section \ref{sec:result}, we show the results including the dependence on
various parameters that control the evolution of grain size distribution.
In Section \ref{sec:discussion}, we provide extended discussion and
additional parameter dependence.
Finally we give conclusions in Section \ref{sec:conclusion}.
For the reference value of the CO-to-H$_2$ conversion factor,
we adopt the MW value as
$X_\mathrm{CO}=2\times 10^{20}$ cm$^{-2}$ K$^{-1}$ km$^{-1}$ s,
which corresponds to
$\alpha_\mathrm{CO}=4.3$ M$_{\sun}$ K$^{-1}$ km$^{-1}$ s pc$^2$
\citep{Bolatto:2013aa}.
We use the solar metallicity $Z_{\sun}$ = 0.014
and the solar oxygen abundance $12+\log\mathrm{(O/H)}_{\sun}=8.7$
\citep{Asplund:2009aa}.

\section{Model}\label{sec:model}

We first review the dust evolution model that incorporates the grain size distribution.
Using the computed grain size distributions at various ages,
we calculate the abundances of H$_2$ and CO in a single typical
dense cloud in the galaxy.
The formation models of these molecules are based on
\citetalias{Hirashita:2017aa}, but are modified to treat the full grain size distribution.
We also predict the CO-to-H$_2$ conversion factor, which is to be compared with observations.
We neglect the spatial structures of the galaxy and the cloud for simplicity and
concentrate on the dependence on the grain size distribution.

\subsection{Evolution of grain size distribution}\label{subsec:review}

The model we adopt for the evolution of grain size distribution in a galaxy is based on
\citetalias{Hirashita:2020aa}, originally developed by \citet{Asano:2013aa} and \citet{Hirashita:2019aa}.
We only provide a summary and refer the interested reader to these papers
for further details.
We also modify the model as described at the end of this
subsection.

We consider two grain species: silicate and carbonaceous dust.
The grain size distribution, %of a grain species $i$,
denoted as $n(a)$, is defined such that
$n(a)\,\mathrm{d}a$ is the number density of grains
at grain radius $a$ within a bin width of $\mathrm{d}a$. The grain radius is related to the
grain mass $m$ as $m=(4\upi /3)a^3s$, where $s$ is the grain material density %%of species $i$
(we adopt $s=3.5$ and 2.24 g cm$^{-3}$ for silicate and carbonaceous dust, respectively;
\citealt{Weingartner:2001aa}).
The grain size distribution is discretized with 128 logarithmic bins in the range of
$a=3$~\AA--10~$\micron$, and adopt $n=0$ at the minimum and maximum grain radii
for the boundary condition.

The galaxy is treated as a one-zone closed box and its chemical evolution is calculated
under a Chabrier initial mass function \citep{Chabrier:2003aa}
and an exponentially decaying SFR with a time-scale of $\tau_\mathrm{SF}$.
%%with a stellar mass range of 0.1--100 M$_{\sun}$.
%%We adopt $\tau_\mathrm{SF}=5$ Gyr following \citetalias{Hirashita:2020aa}
%%for the fiducial model but vary it later.
The main outputs of the chemical evolution model are
the metallicity ($Z$),
the mass abundances of silicon and carbon ($Z_\mathrm{Si}$ and $Z_\mathrm{C}$, respctively),
and the stellar dust production rate
as a function of age $t$.
%%The increment of the metallicity at each time-step is used to calculate the increase of
%%the dust abundance by assuming a condensation efficiency of 10 per cent.
The silicon and carbon abundances
are used to determine the fractions of silicate and carbonaceous dust.
The stellar dust is distributed following a lognormal grain size distribution
centred at $a=0.1~\micron$ with a standard deviation of 0.47.

We also consider the evolution of grain size distribution by the following processes
in the ISM:
dust destruction by supernova (SN) shocks, dust growth by the accretion of
gas-phase metals in the dense ISM, grain growth (sticking) by coagulation
in the dense ISM and grain fragmentation/disruption by
shattering in the diffuse ISM. These processes are simply referred to as
SN destruction, accretion, coagulation, and shattering, respectively.
We fix the mass fraction of the dense ISM,
$\eta_\mathrm{dense}$, and adopt
$(n_\mathrm{H}/\mathrm{cm}^{-3},\, T_\mathrm{gas}/\mathrm{K})=(0.3,\, 10^4)$
and $(300,\, 25)$ for the diffuse and dense ISM, respectively \citep[see also][]{Yan:2004aa},
where $n_\mathrm{H}$ is the hydrogen number density and $T_\mathrm{gas}$ is the gas temperature.
We treat $\eta_\mathrm{dense}$ as a constant parameter for simplicity.
Coagulation and shattering are particularly important to redistribute the grains in large and small grain
radii, respectively, contributing to realizing a smooth power-law-like grain size distribution.
%%The evolution of $n_i$ through coagulation and shattering is treated by
%%the Smolukovski equation (or its modification)
Since coagulation and shattering occur exclusively in the dense and diffuse ISM, respectively,
we calculate these processes with weighting factors of $\eta_\mathrm{dense}$ and
$(1-\eta_\mathrm{dense})$, respectively.
Dust growth by accretion occurs only in the dense ISM, so that the weighting factor
$\eta_\mathrm{dense}$
is also applied to this process. Accretion plays an important role in increasing the grain abundance
at intermediate and late epochs.
We also include SN destruction, which is assumed to occur in both
ISM phases.
{The efficiency of SN destruction is uncertain and dependent on
pre-SN density structure of the ambient ISM
\citep[e.g.][]{Priestley:2021aa} and on detailed dust processing (e.g.\ shattering)
associated with SN shocks \citep[e.g.][]{Jones:1996aa,Kirchschlager:2022aa}.
However, this uncertainty does not have a large impact on our conclusions since
the evolution of grain size distribution is, in our model, predominantly
driven by the other processes mentioned above \citep{Hirashita:2019aa}.}
At each time-step, we calculate
the dust-to-gas ratio, $\mathcal{D}$, by integrating the
grain size distribution weighted with the grain mass and divided by the gas mass density,

There are two modifications applied to the \citetalias{Hirashita:2020aa} model.
Since the original model overestimates the dust-to-metal ratio, we impose the upper limit
for it; that is, we set a maximum of $\mathcal{D}=\mathrm{(D/Z)_{max}}Z$, and
adopt $\mathrm{(D/Z)_\mathrm{max}}=0.48$ following \citet{Hirashita:2023aa}.
At $Z\sim 1~\mathrm{Z}_{\sun}$, the dust-to-metal ratio approaches
$\mathrm{(D/Z)_{max}}$, which is consistent with the value observed
in nearby solar-metallicity galaxies \citep[e.g.][]{Clark:2016aa,Chiang:2021aa}.
The other modification is regarding
the treatment of the dust species, also following
\citet{Hirashita:2023aa}.\footnote{\citet{Hirashita:2023aa} also modified the treatment of
interstellar processing for small carbonaceous grains, but this modification is not included in this
paper. Although we may underestimate the grain surface area,
the H$_2$ formation is already efficient before the age when this modification becomes important.
The extinction, which is also important for CO formation, is little affected by
\citet{Hirashita:2023aa}'s treatment. Thus, neglecting this modification does not
affect our results.}
Since our model is not capable of treating interspecies interaction,
we calculate the evolution of grain size distribution twice by assuming all grains are silicate firstly,
and graphite secondly. We later multiply the silicate grain size distribution by the silicate mass fraction
$f_\mathrm{sil}$ and the carbonaceous grain size distribution by $1-f_\mathrm{sil}$.
The silicate mass fraction is calculated by $f_\mathrm{sil}=6Z_\mathrm{Si}/(6Z_\mathrm{Si}+Z_\mathrm{C})$
at each age, where the factor 6 accounts for the mass fraction of silicon in silicate.
%%The grain size distributions of silicate and carbonaceous dust are obtained by multiplying
%%$n(a)$ in each calculation by $f_\mathrm{sil}$ and $1-f_\mathrm{sil}$, respectively.
Nevertheless, the grain size distributions are not sensitive to the material properties.
%%, which affect the efficiencies of interstellar processing.
Thus, the obtained grain size distributions are
almost identical to those obtained by \citetalias{Hirashita:2020aa}.
The carbonaceous component is further divided into
aromatic and non-aromatic components according to the aromatic fraction at each grain
radius.
The aromatic fraction is approximately equal to $1-\eta_\mathrm{dense}$.
The finally obtained grain size distributions are denoted as
$n_\mathrm{sil}(a)$, $n_\text{ar}(a)$ and $n_\text{non-ar}(a)$ for silicate, aromatic, and
non-aromatic grains, respectively.

\subsection{H$_2$ abundance}\label{subsec:H2}

We consider a typical dense cloud that has similar gas density and temperature to
those in `molecular clouds' in the MW environment.
The hydrogen number density and the gas temperature in this cloud are denoted as
$n_\mathrm{H,cl}$ and $T_\mathrm{cl}$, respectively.
Note that this cloud is not necessarily fully molecular at low metallicity.
{The reaction rates are evaluated under an assumption that the density and dust-to-gas
ratio are uniform in the cloud.
Possible impacts of inhomogeneity is discussed later in Section \ref{subsec:inhomogeneity}.}
By choosing the physical conditions similar to those in the MW,
we are able to examine if this cloud has a conversion factor similar to the MW value at
solar metallicity.
We apply the grain size distribution at each epoch (metallicity) calculated by the
method in Section \ref{subsec:review}.
This implicitly assumes that the grain size distribution is the same in any part (or gas phase)
of the galaxy. This assumption is just due to our one-zone treatment of the galaxy,
but a possible future improvement is given in Section \ref{subsec:prescription}.

\citetalias{Hirashita:2017aa} calculated the H$_2$ abundance
under the two-size approximation.
We extend this to the full treatment of grain
size distribution.
We consider the H$_2$ abundance in a cloud
(`\textit{typical cloud}') with a typical hydrogen column density of
$N_\mathrm{H}(\sim 10^{22}~\mathrm{cm}^{-2})$.
The H$_2$ formation rate is proportional to
the local density represented by the
number density of hydrogen nuclei $n_\mathrm{H,cl}(\sim 10^3~\mathrm{cm}^{-3})$.
For simplicity, we treat the cloud as a one-zone object; thus,
we assume that $n_\mathrm{H,cl}$ is constant and the shielding of dissociating
radiation is given by the column of $N_\mathrm{H}$.
We refer the interested reader to \citet{Krumholz:2008aa,Krumholz:2009aa}
for a spatially resolved approach.

We assume that the H$_2$ abundance is determined by the equilibrium
between the formation on dust surfaces and the dissociation by the interstellar radiation
field (ISRF).
{We neglect formation of H$_2$ through gas-phase reactions, whose effects are
commented at the end of this subsection.}
We define the H$_2$ fraction, $f_\mathrm{H_2}$, as the fraction of hydrogen
nuclei in the form of molecular hydrogen. With this definition,
the increasing rate of $f_\mathrm{H_2}$ by H$_2$ formation on grain surfaces
is evaluated as \citep{Yamasawa:2011aa}
\begin{align}
\left[\frac{\mathrm{d}f_\mathrm{H_2}}{\mathrm{d}t}\right]_\mathrm{form}=
\sum_i(1-f_\mathrm{H_2})S_\mathrm{H}\bar{v}\int_0^\infty\upi a^2n_i(a)\,\mathrm{d}a,
\label{eq:formation}
\end{align}
where the summation is taken for the grain species $i$ ($i=\mathrm{sil}$, ar, and non-ar),
$S_\mathrm{H}$ is the probability that a hydrogen atom incident on the dust surface
reacts with another hydrogen atom to form H$_2$, 
$\bar{v}$ is the mean thermal speed, and
$m_\mathrm{H}$ is the atomic mass of hydrogen.
We fix $S_\mathrm{H}=0.3$ for all the grain species: such a high value is appropriate
in cold and shielded environments \citep{Hollenbach:1979aa}.
The thermal speed $\bar{v}$ is evaluated
as \citep{Spitzer:1978aa}
\begin{align}
\bar{v}=\sqrt{\frac{8k_\mathrm{B}T_\mathrm{cl}}{\pi m_\mathrm{H}}},
\end{align}
where $k_\mathrm{B}$ is the Boltzmann constant.
%%We adopt $T_\mathrm{gas}=25$ K, which is also adopted for the dense phase
%%in the dust evolution calculations (Section \ref{subsec:review}).

For the dissociation of H$_2$, 
the changing rate of the molecular fraction is estimated using the
rate coefficient $R_\mathrm{diss}$ as
\begin{align}
\left[\frac{\mathrm{d}f_\mathrm{H_2}}{\mathrm{d}t}\right]_\mathrm{diss}
=-R_\mathrm{diss}f_\mathrm{H_2}.\label{eq:diss}
\end{align}
The rate coefficient is given by
\citep{Hirashita:2005aa}
\begin{align}
R_\mathrm{diss}=4.4\times 10^{-11}\chi S_\mathrm{shield,H_2}
S_\mathrm{shield,dust}~\mathrm{s}^{-1},
\end{align}
where the factors $S_\mathrm{shield,H_2}$
and $S_\mathrm{shield,dust} (\leq 1)$ are the suppression factors
by H$_2$ self-shielding and dust
extinction, respectively. We adopt the following form for $S_\mathrm{shield}$
\citep{Draine:1996aa,Hirashita:2005aa}:
\begin{align}
S_\mathrm{shield,H_2}=\min\left[1,\,\left(
\frac{\frac{1}{2}f_\mathrm{H_2}N_\mathrm{H}}{10^{14}~\mathrm{cm}^{-2}}
\right)^{-0.75}\right] ,
\end{align}
and
\begin{align}
S_\mathrm{shield,dust}=\exp\left(-\sum_i\tau_{\mathrm{LW},i}\right) ,
\label{eq:shield_dust}
\end{align}
where $\tau_{\mathrm{LW},i}$ is the optical depth of
dust component $i$ at the Lyman-Werner (LW) band, and $\chi$ is
the UV radiation field intensity at the LW
band normalized to the solar neighbourhood value
derived by \citet{Habing:1968aa},
$3.2\times 10^{-20}$ erg s$^{-1}$ cm$^{-2}$ Hz$^{-1}$ sr$^{-1}$;
see also \citet{Hirashita:2005aa}.
Note that $\chi =1.7$ corresponds to the Galactic radiation field derived by \citet{Draine:1978aa}.

The optical depth $\tau_{\mathrm{LW},i}$ is
estimated using Mie theory \citep{Bohren:1983aa} based on
the grain size distributions calculated in Section \ref{subsec:review}.
The optical properties of silicate, aromatic, and non-aromatic components are
taken from astronomical silicate, graphite, and amorphous carbon, respectively.
The first two species are the same as those adopted by \citet{Weingartner:2001aa},
and the last one is taken from the `ACAR' in \citet{Zubko:1996aa}, following
\citetalias{Hirashita:2020aa}.
This calculation outputs the dust extinction optical depth per hydrogen nucleus
at a representative
wavelength for the LW band (1000~\AA), which is multiplied by $N_\mathrm{H}$
to obtain $\tau_{\mathrm{LW},i}$ for each species.
%%The extinction curve becomes similar to the MW curve at $t\sim 10$ Gyr
%%\citepalias{Hirashita:2020aa}.

We treat the UV field $\chi$ as a fixed parameter. In reality, the UV field is
related to the SFR of the galaxy, but the relation depends on various factors
such as the distribution of stars and dust, the extinction in the diffuse ISM, etc.
To avoid including more complex assumptions, we simply treat $\chi$ as a
free parameter and separately examine the dependence on $\chi$ (Section \ref{subsec:others}).

We finally obtain $f_\mathrm{H_2}$ by assuming the equilibrium condition:
$[\mathrm{d}f_\mathrm{H_2}/\mathrm{d}t]_\mathrm{form}+
[\mathrm{d}f_\mathrm{H_2}/\mathrm{d}t]_\mathrm{diss}=0$, which is evaluated
using equations (\ref{eq:formation}) and (\ref{eq:diss}).
As discussed in \citetalias{Hirashita:2017aa}, the equilibrium assumption is
reasonable if the dust-to-gas ratio is larger than $\sim 10^{-3}$, roughly corresponding
to $Z\gtrsim 0.1$ Z$_{\sun}$ (Section \ref{subsec:comparison}).
Indeed, \citet{Hu:2023aa}, based on a hydrodynamic simulation focused on
a dwarf galaxy with $Z=0.1$ Z$_{\sun}$, showed that the H$_2$ abundance could be suppressed
because of a long H$_2$ formation time \citep[see also][]{Hu:2021aa}.
{We also note that, in the metallicity range where the equilibrium holds,
the results are insensitive to other sources of molecules such as stellar ejecta,
unless the supply occurs quickly in dense clouds.
Formation of H$_2$ in the gas phase 
\citep[as listed in e.g.][]{Galli:1998aa,Hirata:2006aa} is usually
negligible \citep[e.g.][]{Hirashita:2002aa} as well in the above metallicity range.
In particular, the gas-phase formation is only
able to raise $f_\mathrm{H_2}$ up to $\sim 10^{-3}$--$10^{-2}$
\citep[e.g.][]{Romano:2022aa}.
These molecular sources and formation paths that are not included in this paper
could raise the H$_2$ (and also CO) abundances at
low metallicity, especially at $Z<0.1$ Z$_{\sun}$.
In particular, if we are interested in the regime where $f_\mathrm{H_2}$ is
smaller than $\sim 10^{-2}$, there is a risk of underestimating $f_\mathrm{H_2}$
because we neglect the H$_2$ formation in the gas phase. Thus,
for direct comparison with observational data, we
only focus on $Z>0.1$ Z$_{\sun}$. We, nevertheless, expect that theoretical
predictions at lower metallicity still give useful qualitative insights into
how the molecular fraction decreases with decreasing metallicity.}
An Implementation of our grain size evolution model into a hydrodynamic
simulation is needed to fully understand nonequilibrium effects on the H$_2$
abundance at low metallicity, which is left for future work.

\subsection{CO abundance}\label{subsec:fco}

We utilize the CO abundance calculations for various
physical conditions by \citet[][hereafter GM11]{Glover:2011aa}, who computed
H$_2$ and CO abundances in 5--20 pc boxes
using hydrodynamic simulations coupled with
chemical network calculations for H$_2$ and CO (see also \citealt{Shetty:2011aa}).
{Note that this effectively takes into account the inhomogeneity in gas density.}
\citetalias{Hirashita:2017aa} used \citet{Feldmann:2012aa}'s method to
interpolate or extrapolate reasonably the calculated data, since
running new time-consuming chemical network calculations
for various metallicities and grain size distributions is not realistic for our work.

The CO abundance $f_\mathrm{CO}$ (denoted as $x_\mathrm{CO}$ in
\citetalias{Hirashita:2017aa}) is defined as the ratio of
CO molecules to hydrogen nuclei in number.
We assume that the CO abundance is determined by
$A_V$ ($V$-band extinction used as an indicator of dust extinction), $\chi$, and $Z$:
$f_\mathrm{CO} = f_\mathrm{CO}(A_V, \chi, Z)$.
%%We impose an upper limit for $f_\mathrm{CO}$ by the carbon abundance $f_\mathrm{C} = 1.41 \times 10^{-4} Z/Z_{\odot}$.
The basic idea in \citet{Feldmann:2012aa} is to find an extinction value
$A'_V$ in \citetalias{Glover:2011aa}'s system that satisfies
$f'_\mathrm{CO}(A'_V,\,\chi '=1.7,\, Z)=f_\mathrm{CO}(A_V,\,\chi,\,Z)$,
where the notations with a prime indicate the values in \citetalias{Glover:2011aa}'s system.
We compare the two systems at the same metallicity, so that $Z^{\prime} = Z$.
Note that \citetalias{Glover:2011aa} calculated CO fraction $f'_\mathrm{CO}$ with $\chi'=1.7$.
The following two fitting formulae hold for the quantities in \citetalias{Glover:2011aa}:
\begin{equation}
    f^{\prime}_\mathrm{H_{2}} = 1 - \exp(-0.45A^{\prime}_V),
	\label{eq:fH2prime}
\end{equation}
and 
\begin{equation}
    \log_{10} f^{\prime}_\mathrm{CO} = -7.64 + 3.89 \log_{10} A^{\prime}_V.
	\label{eq:xCOprime}
\end{equation}
The following relation also holds in \citetalias{Glover:2011aa}'s system:
\begin{equation}
    N^{\prime}_\mathrm{H} = \frac{A^{\prime}_V}{5.348 \times 10^{-22} (Z^{\prime} / Z_{\odot})~\mathrm{cm}^{2}}.
	\label{eq:NHprime}
\end{equation}

\citet{Feldmann:2012aa} assume that, if $f_\mathrm{CO}=f'_\mathrm{CO}$,
the CO formation rate (and the CO dissociation rate in equilibrium)
is comparable in the two systems. Thus, we search for a condition in which
the dissociation rate is the same. This condition is written as
(note that
$f_\mathrm{CO} = f^{\prime}_\mathrm{CO}$, so $N_\mathrm{CO} = f^{\prime}_\mathrm{CO} N_\mathrm{H}$, where $N_\mathrm{CO}$ is the CO column density) 
\begin{equation}
\begin{aligned}
	& {\chi}S_\mathrm{dust}(A_{V\mathrm{,eff}})S_\mathrm{H_2}(f_\mathrm{H_2} N_\mathrm{H} / 2)S_\mathrm{CO}(f^{\prime}_\mathrm{CO}N_\mathrm{H})/N_\mathrm{H} \\
	& = 1.7S_\mathrm{dust}(A^{\prime}_V)S_\mathrm{H_{2}}(f^{\prime}_\mathrm{H_{2}} N^{\prime}_\mathrm{H} / 2)S_\mathrm{CO}(f^{\prime}_\mathrm{CO}N^{\prime}_\mathrm{H})/N^{\prime}_\mathrm{H},
	\label{eq:xCOre}
\end{aligned}
\end{equation}
where $S_\mathrm{dust}(A_{V,\mathrm{eff}})$,
$S_\mathrm{H_2}(f_\mathrm{H_2} N_\mathrm{H} / 2)$, and
$S_\mathrm{CO}(x^{\prime}_\mathrm{CO}N_\mathrm{H})$ are
the shielding factors of CO-dissociating photons by dust, H$_2$, and CO, respectively,
taken from \citet{Lee:1996aa}, and
$A_{V,\mathrm{eff}} = A_{1000~\angstrom}/4.7$ is the effective $V$-band extinction,
of which the relation to the value at 1000~\AA\ is obtained from the MW extinction curve
since the MW dust
properties are implicitly assumed in \citetalias{Glover:2011aa}'s system.
We evaluate $A_{1000~\angstrom} \simeq 1.086 \sum_{i}^{} \tau_{\mathrm{LW},i}$.
Note that $f_\mathrm{H_2}$ is calculated in Section \ref{subsec:H2} and $N_\mathrm{H}$
is a free parameter. With equations (\ref{eq:fH2prime})--(\ref{eq:NHprime}),
$f^{\prime}_\mathrm{H_2}$, $f^{\prime}_\mathrm{CO}$, and $N^{\prime}_\mathrm{H}$ are
written as functions of $A^{\prime}_V$.
Equation~(\ref{eq:xCOre}) is thus
solved for $A^{\prime}_{V}$. Recalling that  $f_\mathrm{CO} = f^{\prime}_\mathrm{CO}$,
we convert the obtained $A^{\prime}_{V}$ to $f_\mathrm{CO}$ using
equation~(\ref{eq:xCOprime}).

We do not constrain the total carbon abundance, since our model is not capable of
reproducing the detailed solar abundance pattern. Although our model includes
major stellar dust and metal sources (core-collapse supernovae and
asymptotic giant branch stars), we do not include other
chemical enrichment sources that do not contribute to the dust production.
This could underestimate some metal elements. However, we confirmed that
the carbon atoms used for CO is less than the half of those locked up in the dust phase.
In our treatment, dust only uses about half of the metals
(Section \ref{subsec:gsd}), so that a minor fraction of the carbon is in the form of CO.
Therefore, we simply let CO form as much as predicted in the above framework.
The detailed chemical abundance treatment, including the metallicity pattern, is left for future work,
and an example of an element-to-element treatment of metal depletion can be seen in e.g.\
\citet{Choban:2022aa}.

\subsection{CO-to-H$_2$ conversion factor}

Based on the above calculations, we estimate the CO-to-H$_2$ conversion
factor as
\begin{align}
X_\mathrm{CO}=N_\mathrm{H_2}/W_\mathrm{CO},\label{eq:Xco}
\end{align}
where $N_\mathrm{H_2}=f_\mathrm{H_2}N_\mathrm{H}/2$ is the column density of
H$_2$.
The intensity of CO emission $W_\mathrm{CO}$ is calculated by
the following expression \citepalias[e.g.][]{Glover:2011aa}:
\begin{align}
W_\mathrm{CO}=T_\mathrm{r}\Delta v\int_0^{\tau_{10}}
2\beta (\tau )\,\mathrm{d}\tau,\label{eq:W_CO}
\end{align}
where $T_\mathrm{r}$ is the observed radiation temperature
(equation \ref{eq:temp_rad}),
$\Delta v$ is the CO line velocity width,
$\tau_{10}$ is the optical depth of the CO $J=1\to 0$ transition,
and $\beta (\tau )$ is the escape probability at optical depth $\tau$.
The last two quantities are given by
\citep{Tielens:2005aa,Feldmann:2012aa}
\begin{align}
\beta (\tau )= \begin{cases}
[1-\exp (-2.34\tau )]/(4.68\tau ) & \mbox{if $\tau\leq 7$}; \\
1/(4\tau [\ln (\tau /\sqrt{\pi})]^{1/2}) & \mbox{if $\tau >7$}.
\end{cases}
\end{align}
and
\begin{align}
\tau_{10}=1.4\times 10^{-16}(1-\mathrm{e}^{-5.5/T_\mathrm{cl}})
\left(\frac{\Delta v}{3~\mathrm{km~s}^{-1}}\right)^{-1}
\left(\frac{N_\mathrm{CO}}{\mathrm{cm}^{-2}}\right) .
\end{align}
The radiation temperature of the CO $J=1\to 0$ transition
is calculated by
\begin{align}
T_\mathrm{r}=5.5\left(\frac{1}{\mathrm{e}^{5.5/T_\mathrm{cl}}-1}-
\frac{1}{\mathrm{e}^{5.5/T_\mathrm{CMB}}-1}\right)~\mathrm{K} ,
\label{eq:temp_rad}
\end{align}
where $T_\mathrm{CMB}=2.73(1+z)$ is the CMB temperature
(we adopt redshift $z=0$ in this paper).
Using equation (\ref{eq:W_CO}) together with the H$_2$ column density
in Section \ref{subsec:H2}, we obtain the
$X_\mathrm{CO}$ from equation (\ref{eq:Xco}).

\subsection{Choice of parameter values}\label{subsec:choice}

We need to specify $n_\mathrm{H}$ and $N_\mathrm{H}$ for
the typical cloud.
%%The hydrogen number density $n_\mathrm{H}$ is also necessary in calculating the H$_2$ abundance.
The effects of varying $N_\mathrm{H}$ and $n_\mathrm{H}$ have already been
investigated by \citetalias{Hirashita:2017aa}.
If $N_\mathrm{H}\lesssim 10^{21}$~cm$^{-2}$, the CO abundance is kept low; thus,
CO is hardly detected for such low-column-density clouds.
If $N_\mathrm{H}\gtrsim 10^{23}$~cm$^{-2}$, the CO-to-H$_2$
conversion factor stays $\sim$5 times higher than the MW value at solar metallicity
because the CO emission is optically thick.
Thus, in order to make the prediction consistent with the MW observation,
we adopt $N_\mathrm{H}=10^{22}$~cm$^{-2}$, which is also consistent with
the typical column density of molecular clouds \citep{Solomon:1987aa,Wolfire:2010aa}.
%%This column density corresponds to
%%a surface mass density of 100 M$_{\sun}$ pc$^{-2}$, which is consistent with ...
We still examine later an order of magnitude variation in
$N_\mathrm{H}=0.3$--$3\times 10^{22}$ cm$^{-2}$ (Section \ref{subsec:comparison}).
In the MW condition, this column density corresponds to $A_V\sim 5$, which is consistent with the
region where most of the carbon is in the form of CO
\citep[e.g.][]{Bolatto:2013aa}.
For the number density, we adopt $n_\mathrm{H,cl}=10^3$~cm$^{-2}$, which was also adopted
by \citetalias{Hirashita:2017aa}; however, it is not necessary to assume the density for $f_\mathrm{CO}$
in our formulation since we use the results of \citetalias{Glover:2011aa}, who already adopted typical
densities in their simulation. The above value is consistent with their simulation.
The density still affects $f_\mathrm{H_2}$; however, as we confirm later,
$f_\mathrm{H_2}$ is almost unity in the metallicity range of interest for $X_\mathrm{CO}$.
Therefore, our results for $X_\mathrm{CO}$ is not sensitive to the choice of $n_\mathrm{H,cl}$.
%%In this paper, we concentrate on
%%the effect of the evolution of grain size distribution; thus, we fix
%%$N_\mathrm{H}=10^{22}$ cm$^{-2}$ and $n_\mathrm{H}=300$ cm$^{-3}$.
%%We examine a range of $N_\mathrm{H}=10^{21}$--$10^{23}$ cm$^{-2}$, corresponding
%%to 10--$10^3$ M$_{\sun}$ pc$^{-2}$, which
%%covers the range of surface densities of giant molecular
%%clouds in various environments \citep[e.g.][]{Bolatto:2013aa}.
%%% see also Bolatto 2013 ARA\&A Section 4.4

In addition, we also need $T_\mathrm{cl}$ and $\chi$ for the physical condition of the
typical cloud. We adopt $T_\mathrm{cl}=10$ K for the fiducial value following the
previous studies (\citealt{Feldmann:2012aa}; \citetalias{Hirashita:2017aa}).
At low metallicity, where the line is optically thin, $W_\mathrm{CO}$
is insensitive to $T_\mathrm{cl}$ because higher $T_\mathrm{cl}$ raises
the emissivity but decreases the optical depth.
In contrast, at high metallicity, the system is optically thick, so that
$W_\mathrm{CO}$ is roughly proportional to $T_\mathrm{cl}$.
The velocity dispersion $\Delta v$ has almost the same influence on $W_\mathrm{CO}$ as
$T_\mathrm{cl}$.
%%; thus, $T_\mathrm{cl}$ and $\Delta v$ are degenerate in the calculation of $X_\mathrm{CO}$.
Thus, we fix
$\Delta v=3$ km s$^{-1}$ \citep{Feldmann:2012aa}.
%%, and vary $T_\mathrm{cl}$ in the range of 10--100 K in Section \ref{subsec:others}.
For $\chi$, we assume the Galactic value ($\chi =1.7$) in the
fiducial model. Nevertheless, we still address different values of $T_\mathrm{cl}$ and
$\chi$ later in discussing galaxies in which the physical conditions are very different from those
in the MW (Section \ref{subsec:others}).
More detailed discussions on the dependence on various environmental parameters are
given by \citet{Maloney:1988aa}.

In the dust evolution model, the important parameters are the dense gas fraction
$\eta_\mathrm{dense}$ and the star formation time-scale $\tau_\mathrm{SF}$.
The first parameter $\eta_\mathrm{dense}$ affects the
functional shape of the grain size distribution mainly through the balance between
shattering and coagulation (Section \ref{subsec:gsd}).
%%: a larger value of $\eta_\mathrm{dense}$ 
%%predicts a grain size distribution more dominated by larger grains because of
%%more efficient coagulation, which occurs in the dense ISM. A smaller $\eta_\mathrm{dense}$
%%leads to more efficient shattering, which takes place in the diffuse ISM; as a consequence,
%%the grain size distribution is more biased to smaller sizes.
Therefore, the variation of $\eta_\mathrm{dense}$ serves
to examine the effect of grain size distribution on the molecular abundances.
The second parameter ($\tau_\mathrm{SF}$) regulates the time-scale of
metal (and dust) enrichment. In this paper we examine the parameter values
$\eta_\mathrm{dense}=0.1$, 0.5 (fiducial), and 0.9, and
$\tau_\mathrm{SF}=0.5$, 5 (fiducial), and 50 Gyr.

\section{Results}\label{sec:result}

We show the evolution of $f_\mathrm{H_2}$, $f_\mathrm{CO}$,
and $X_\mathrm{CO}$ for various values of $\eta_\mathrm{dense}$ and
$\tau_\mathrm{SF}$.
We first show the evolution of grain size distribution,
but refer the interested reader to 
\citetalias{Hirashita:2020aa} for detailed discussion.
The main focus in this section is put on the results for molecules.
To present the evolution, we use the metallicity ($Z$) instead of the time ($t$)
since the metallicity is easier to obtain observationally.
In our exponentially decaying star formation history,
$Z/\mathrm{Z}_{\sun}\simeq 1.3(t/\tau_\mathrm{SF})$ approximately holds
between 0.01 to 1 Z$_{\sun}$.
%%if the reader needs to convert $Z$ to $t$.

\subsection{Evolution of grain size distribution}\label{subsec:gsd}

\begin{figure}
 \includegraphics[width=\columnwidth]{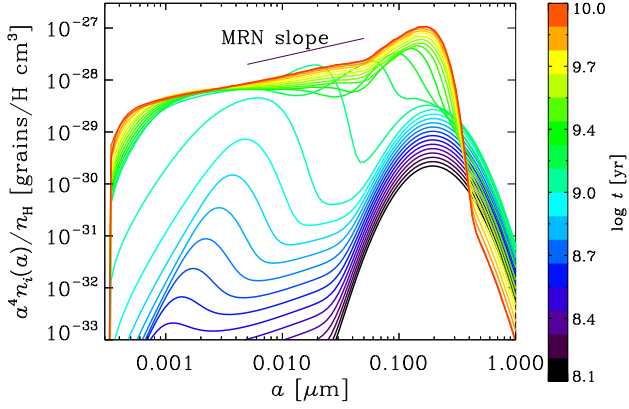}
 \caption{Evolution of grain size distribution. We present the case for silicate, but a similar
 evolutionary sequence is also obtained for carbonaceous dust. The grain size distribution
 is multiplied by $a^4$ and divided by $n_\mathrm{H}$ so that the resulting quantity is
 proportional to the dust mass in logarithmic bins per gas mass. The colour indicates the
 age as shown in the colour bar. The slope of the \citetalias{Mathis:1977aa} grain size distribution
 is shown by the thin solid line for reference.}
 \label{fig:gsd}
\end{figure}

For the convenience in interpreting the results below, we present
the evolution of grain size distribution for
the fiducial case ($\eta_\mathrm{dense}=0.5$ and $\tau_\mathrm{SF}=5$~Gyr)
in Fig.~\ref{fig:gsd} (see \citetalias{Hirashita:2020aa} for detailed discussion).
Since silicate and carbonaceous dust have similar evolutionary trends in the
grain size distribution, we only show the results for silicate.
In the early epoch ($t\lesssim 0.3$~Gyr), large ($a\sim 0.1~\micron$) grains,
which are supplied from stellar sources,
dominate the overall grain population. After that,
shattering gradually produces a tail of the grain size distribution extending
towards small radii. At $t\sim 1$ Gyr, accretion causes a drastic increase of
small grains because of their large surface area. At ages greater than a few Gyr,
coagulation creates large grains from small grains, forming a smooth grains size
distribution. At the same time, shattering continues to disrupt large grains, determining
the upper cut-off of grain radius at $a\sim 0.2~\micron$.
The grain size distribution at $t\sim 10$ Gyr eventually becomes a shape similar to
the \citetalias{Mathis:1977aa} grain size distribution $n\propto a^{-3.5}$, and
the functional shape is determined
by the balance between coagulation and shattering
\citep[see also][]{Dohnanyi:1969aa,Tanaka:1996aa,Kobayashi:2010aa}.

The effects of $\eta_\mathrm{dense}$ and $\tau_\mathrm{SF}$ are also shown and
discussed in \citetalias{Hirashita:2020aa}. They are summarized as follows.
If $\eta_\mathrm{dense}$ is smaller, coagulation becomes less efficient so that
small grains less efficiently stick to form large grains.
As a consequence,
the grain size distributions
at later times ($t\sim 3$--10~Gyr) are more dominated by small grains for smaller
$\eta_\mathrm{dense}$. The opposite trend is observed for larger $\eta_\mathrm{dense}$;
that is, the grain size distribution extends to larger grain radii.
If we adopt $\eta_\mathrm{dense}=0.1$ (0.9), the upper cut-off of grain radius
is located at $a\sim 0.08$ (0.8) $\micron$ for silicate;
0.04 (0.8) $\micron$ for carbonaceous dust.
The other parameter, $\tau_\mathrm{SF}$, effectively regulates the speed of
metal enrichment by stars.
Faster enrichment overall leads to quicker dust evolution;
however, since the time-scale of interstellar processing of dust
does not scale with $\tau_\mathrm{SF}$ under a given metallicity,
quicker metal enrichment means that the rate of interstellar processing
catches up with that of metal enrichment at higher metallicity.
This is most clearly seen in the evolution of dust-to-gas ratio as we will present
in Section \ref{subsec:tau_SF}.
As mentioned in \citetalias{Hirashita:2020aa} (see also \citealt{Asano:2013ab}),
a similar functional shape of grain size distribution is realized at the same value of
$t/\tau_\mathrm{SF}^{1/2}$: this means that, since $Z\propto t/\tau_\mathrm{SF}$ approximately holds,
a similar grain size distribution is realized at the same value of $Z\tau_\mathrm{SF}^{1/2}$, confirming
the above statement that the modification of grain size distribution by interstellar processing
occurs at higher metallicity for shorter $\tau_\mathrm{SF}$.

\subsection{Dependence on $\eta_\mathrm{dense}$}\label{subsec:eta}

\begin{figure}
 \includegraphics[width=\columnwidth]{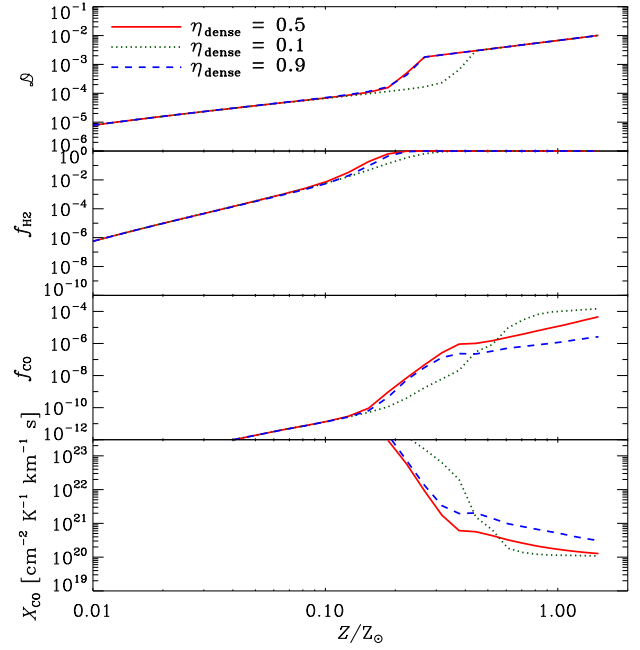}
 \caption{
 Panels from upper to lower show the dust-to-gas ratio ($\mathcal{D}$),
 H$_2$ fraction ($f_\mathrm{H_2}$), CO fraction ($f_\mathrm{CO}$)
 and CO-to-H$_2$ conversion factor ($X_\mathrm{CO}$) as functions of metallicity ($Z$).
 The solid, dotted, and dashed lines show the results for $\eta_\mathrm{dense}=0.5$
 (fiducial), 0.1, and 0.9, respectively.
 $X_\mathrm{CO}$ is presented only when $f_\mathrm{CO}>10^{-10}$.
 For parameters other than $\eta_\mathrm{dense}$, we adopt the fiducial values.}
 \label{fig:mol_eta}
\end{figure}

We investigate the evolution of H$_2$ and CO abundances
under various values of $\eta_\mathrm{dense}$, which regulates the
grain size distribution. We adopt the fiducial values for the parameters
of the cloud properties ($N_\mathrm{H}=10^{22}$~cm$^{-3}$,
$\chi =1.7$, $n_\mathrm{H,cl}=10^3$ cm$^{-2}$,
$T_\mathrm{cl}=10$~K, and $\Delta v=3$ km s$^{-1}$).
In Fig.\ \ref{fig:mol_eta}, we show the metallicity dependence of the
molecular abundances and the CO-to-H$_2$ conversion factor.
$X_\mathrm{CO}$ is only shown where $f_\mathrm{CO}>10^{-10}$,
since it is not meaningful to show $X_\mathrm{CO}$ below such a low
undetectable CO abundance.
%%(i.e.\ before the CO abundance is significantly increased by the growth of dust abundance).
These quantities are shown as a function of
metallicity, which is used as an indicator of galaxy evolution.

The evolution of $\mathcal{D}$ is affected by $\eta_\mathrm{dense}$
(the first panel in Fig.~\ref{fig:mol_eta}).
In particular, the metallicity at which $\mathcal{D}$ steeply increases depends
on $\eta_\mathrm{dense}$:
$Z\sim 0.2$--0.3~Z$_{\sun}$ for $\eta_\mathrm{dense}=0.5$ and 0.9, and
$Z\sim 0.3$--0.4~Z$_{\sun}$ for $\eta_\mathrm{dense}=0.1$.
This steep increase of $\mathcal{D}$ is due to dust growth by accretion.
Below this metallicity,
the dust is predominantly supplied by stars. If $\eta_\mathrm{dense}$ is as small as
0.1, the fraction of the dense ISM, which hosts accretion, is small.
This leads to a low efficiency of accretion, delaying the increase of $\mathcal{D}$.
In this context, the case with $\eta_\mathrm{dense}=0.9$ should show the most
efficient dust growth; however,
the accretion of gas-phase metals most efficiently occurs for
small grains, whose production by shattering is the most inefficient
in the case of the largest $\eta_\mathrm{dense}$ (because of the lowest fraction
of the diffuse ISM hosting shattering).
Because of these two counteracting effects, the metallicity at which $\mathcal{D}$
rapidly increases is not different between the cases with $\eta_\mathrm{dense}=0.9$
and 0.5. At low and high $Z$, $\mathcal{D}$ does not depend on $\eta_\mathrm{dense}$
because at low $Z$, the stellar dust production, which is independent of $\eta_\mathrm{dense}$,
dominates the dust mass increase and at high $Z$, the dust-to-metal ratio is saturated to the
maximum value $\mathrm{(D/Z)_{max}}$.

The H$_2$ fraction ($f_\mathrm{H_2}$) in the typical dense cloud also increases with metallicity
(the second panel in Fig.~\ref{fig:mol_eta}).
In the fiducial case ($\eta_\mathrm{dense}=0.5$),
the increase of $f_\mathrm{H_2}$ is accelerated at $Z\gtrsim 0.1$ Z$_{\sun}$ and
$f_\mathrm{H_2}$ approaches 1 at $Z\sim 0.2$ Z$_{\sun}$.
At $Z\sim 0.1$ Z$_{\sun}$ ($t\sim 0.7$ Gyr), the abundance of small grains
starts to increase significantly by shattering and accretion (Fig.\ \ref{fig:gsd}).
This accelerates the
increase in the surface area, raising the H$_2$ formation rate.
Self-shielding of H$_2$ further increases the H$_2$ abundance by suppressing the
H$_2$ dissociation.
Although dust shielding also increases in this phase,
it does not play a significant role in increasing the H$_2$ abundance. 
Indeed,
%%by the calculation without dust extinction 
we confirm (not shown in the figure) that there is
little difference in $f_\mathrm{H_2}$ between the cases with and without dust shielding
(i.e.\ applying $\tau_{\mathrm{LW},i}=0$ for the latter).
The importance of self-shielding is also addressed by a hydrodynamic simulation
which included the evolution of both grain size distribution and H$_2$ formation
\citep{Romano:2022aa}.
The H$_2$ fraction also depends on $\eta_\mathrm{dense}$: for $\eta_\mathrm{dense}=0.1$,
the increase of small grains by accretion is delayed, so that the
increase of $f_\mathrm{H_2}$ also occurs at a later stage.
Since shattering is slower for $\eta_\mathrm{dense}=0.9$ than for $\eta_\mathrm{dense}=0.5$,
the increase of $f_\mathrm{H_2}$ is slightly delayed for the larger value of $\eta_\mathrm{dense}$.

The CO fraction ($f_\mathrm{CO}$) in the typical dense cloud increases with metallicity
(the third panel in Fig.\ \ref{fig:mol_eta}).
The metallicity at which $f_\mathrm{CO}$ steeply increases
corresponds to the phase in which dust growth by accretion starts to play a significant
role in increasing the dust abundance ($Z\sim 0.2$ Z$_{\sun}$ in the fiducial case).
Thus, dust shielding is important for the increase of $f_\mathrm{CO}$.
In Appendix \ref{app:CO}, we examine the effect of each shielding source.
The results are shown in Fig.\ \ref{fig:CO_appendix}.
If we calculate $f_\mathrm{CO}$ without dust extinction
($A_{V,\mathrm{eff}}=0$), the CO abundance is significantly underpredicted.
H$_2$ shielding plays a minor but appreciable role after the cloud becomes fully molecular,
while CO self-shielding becomes as effective as H$_2$ shielding only after the
increase of $f_\mathrm{CO}$ at $Z\sim 0.3$ Z$_{\sun}$ (Fig.\ \ref{fig:CO_appendix}).

As we observe in Fig.\ \ref{fig:mol_eta}, the $f_\mathrm{CO}$--$Z$ relation
varies with $\eta_\mathrm{dense}$.
In the case of $\eta_\mathrm{dense}=0.1$, the increase of $f_\mathrm{CO}$ occurs at a later
stage because the increase of dust abundance occurs later. At high metallicity,
$f_\mathrm{CO}$ eventually reaches a higher value for $\eta_\mathrm{dense}=0.1$ than
for higher $\eta_\mathrm{dense}$.
This is because the dust extinction is enhanced if the grain size distribution is biased towards smaller
sizes. The grain size effect is more effectively seen if we compare the results for
$\eta_\mathrm{dense}=0.5$ and 0.9. Although the evolution of dust-to-gas ratio is similar between
these two cases, the resulting $f_\mathrm{CO}$ is differentiated because of the difference in the
grain size distribution.
The grain size distribution is more biased to larger grains for $\eta_\mathrm{dense}=0.9$ than for
$\eta_\mathrm{dense}=0.5$, leading to less efficient shielding of dissociating radiation.
Therefore, the evolution of grain size distribution and dust abundance
has a large impact in the metallicity
dependence of $f_\mathrm{CO}$.

As expected from the sensitive hevaviour of $f_\mathrm{CO}$ to the grain evolution,
the CO-to-H$_2$ conversion factor, $X_\mathrm{CO}$, is greatly affected by
$\eta_\mathrm{dense}$ (the bottom panel in Fig.\ \ref{fig:mol_eta}).
First of all, $X_\mathrm{CO}$ is sensitive to metallicity as already shown by
other studies (see the Introduction).
The change of $X_\mathrm{CO}$ particularly occurs in the metallicity range
where the cloud is rich in H$_2$ but is not rich in
CO. $X_\mathrm{CO}$ approaches the Milky Way value
($\sim 2\times 10^{20}$ cm$^{-2}$ K$^{-1}$ km$^{-1}$ s;
\citealt{Bolatto:2013aa}) at nearly solar metallicity.
The decrease of $X_\mathrm{CO}$ towards high
metallicity roughly traces the increasing trend of $f_\mathrm{CO}$.
The different tracks for the different values of $\eta_\mathrm{dense}$ also
follows the trends in $f_\mathrm{CO}^{-1}$.
The conversion factor stays relatively high for $\eta_\mathrm{dense}=0.9$,
while it drops down to $\sim 10^{20}$ cm$^{-2}$ K$^{-1}$ km$^{-1}$ s for the
other cases.

\subsection{Dependence on $\tau_\mathrm{SF}$}\label{subsec:tau_SF}

\begin{figure}
 \includegraphics[width=\columnwidth]{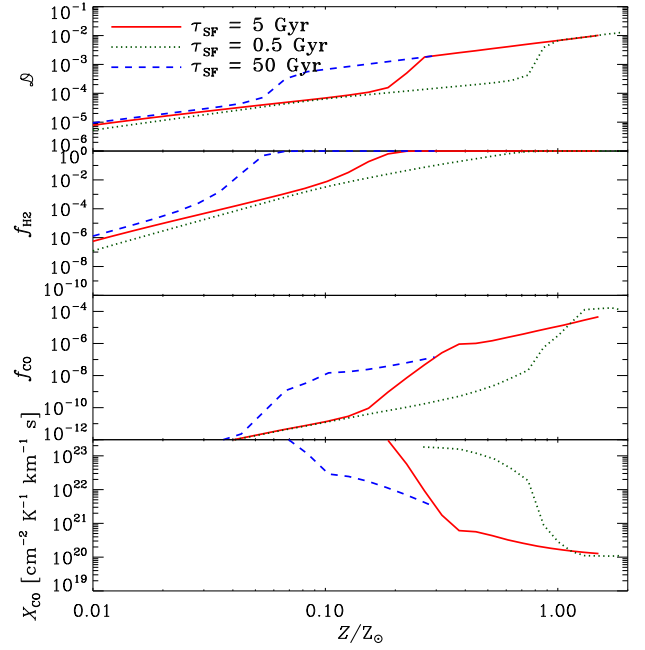}
 \caption{Same as Fig.\ \ref{fig:mol_eta} but for the dependence on
 $\tau_\mathrm{SF}$. We adopt the fiducial values for other parameters.
 The solid dotted, and dashed lines show the results for $\tau_\mathrm{SF}=5$ (fiducial),
 0.5, and 50 Gyr, respectively.}
 \label{fig:mol_tau}
\end{figure}

We examine the dependence on the star formation time-scale $\tau_\mathrm{SF}$, which
regulates the speed of metal enrichment. We show
the resulting metallicity dependences of $\mathcal{D}$, $f_\mathrm{H_2}$,
$f_\mathrm{CO}$ and $X_\mathrm{CO}$ in
Fig.\ \ref{fig:mol_tau}.

Overall, the effect of $\tau_\mathrm{SF}$ is to determine the metallicity level at which
steep increases of $\mathcal{D}$, $f_\mathrm{H_2}$, and $f_\mathrm{CO}$ occur.
This is because, as mentioned in Section \ref{subsec:gsd},
a similar grain size distribution is approximately achieved at the same value of
$Z\tau_\mathrm{SF}^{1/2}$ (Section \ref{subsec:gsd}).
%%This means that the rapid increase of small grains
%%occur at $Z\propto\tau_\mathrm{SF}^{-1/2}$.
%%This $\tau_\mathrm{SF}$ dependence emerges from the fact that,
%%if $\tau_\mathrm{SF}$ is long, the accretion can occur more quickly than metal enrichment at
%%low metallicity.
Thus, the evolution of each quantity is `shifted' towards high metallicity
as $\tau_\mathrm{SF}$ becomes shorter.

Since the drop of $X_\mathrm{CO}$ coincides with the rise of $f_\mathrm{CO}$,
the metallicity level at which $X_\mathrm{CO}$ drops is strongly affected by
$\tau_\mathrm{SF}$. In particular, if $\tau_\mathrm{SF}$ is as short as
0.5 Gyr, $X_\mathrm{CO}$ changes sharply from a large value
to the MW-like one at solar metallicity. This means that it is difficult to detect CO from rapidly
($\tau_\mathrm{SF}\lesssim 0.5$ Gyr) star-forming
galaxies if the metallicity is lower than solar. In nearby starbursts, CO is usually detected
probably because they are already sufficiently metal/dust enriched
in the current or previous star formation episodes.

Here we note that treating $\tau_\mathrm{SF}$ as a completely free parameter
could enhance the effect of $\tau_\mathrm{SF}$ on the $\mathcal{D}$--$Z$
relation. \citetalias{Hirashita:2017aa}
did not treat $\tau_\mathrm{SF}$ as a completely independent parameter but
linked it to the accretion time-scale. This is because both accretion (dust growth) and
star formation occur in the dense ISM. In contrast, our present model gives the
dense gas fraction $\eta_\mathrm{dense}$ (note that accretion efficiency is
weighted with $\eta_\mathrm{dense}$ in our approach; Section \ref{subsec:review})
as a free parameter, and
does not relate it to $\tau_\mathrm{SF}$. These different approaches
produce different results in the following point: in
\citetalias{Hirashita:2017aa}'s treatment, the evolution of $\mathcal{D}$ is hardly
affected by $\tau_\mathrm{SF}$ because of the proportionality between
the time-scales of star formation and accretion. In our model, in contrast,
the $\mathcal{D}$--$Z$ relation is strongly affected by $\tau_\mathrm{SF}$.
However, we should also note that \citetalias{Hirashita:2017aa}'s treatment
needs to introduce another free parameter: star formation efficiency.
%%, which was assumed to be independent of metallicity.
Probably, the realistic situation lies between
these two treatments, and can only be treated in a more `realistic' model
such as hydrodynamic simulations that could predict the formation of dense clouds and star formation
consistently.

\subsection{Comparison with observations}\label{subsec:comparison}

We compare the above calculation results with observations.
In particular, data for the dust-to-gas ratio and
the CO-to-H$_2$ conversion factor are available
for nearby galaxies with
different metallicities. We basically use the same observational data
as adopted by \citetalias{Hirashita:2017aa}, who took the data sets
compiled in \citet{Bolatto:2013aa} and supplemented by
\citet{Cormier:2014aa} for low-metallicity galaxies.
In addition to the $X_\mathrm{CO}$--$Z$ relation,
we also show the $\mathcal{D}$--$Z$ relations, where the data are
taken from \citet{Remy-Ruyer:2014aa}.
We adopt the dust-to-gas ratio estimated with a metallicity-dependent
CO-to-H$_2$ conversion factor, which only has a minor influence on the resulting
$\mathcal{D}$--$Z$ relation.
Recent more elaborate analysis \citep[e.g.][]{Aniano:2020aa,DeVis:2021aa,Galliano:2021aa} shows
similar $\mathcal{D}$--$Z$ relations for nearby galaxies.
To further supplement the data for $X_\mathrm{CO}$ at low metallicity, we also include
the data from \citet{Shi:2016aa}, but do not show a galaxy (DDO70-A)
with $Z<0.1$ Z$_{\sun}$, where the gas mass estimate depends strongly on the
assumption on the relation between dust-to-gas ratio and metallicity.
By excluding this galaxy, we concentrate on the comparison at $Z>0.1$ Z$_{\sun}$
as mentioned in Section \ref{subsec:H2}.
%%However, this galaxy ($Z=0.07$ Z$_{\sun}$) is located at a natural extrapolation
%%of the data at $Z>0.1$ Z$_{\sun}$.

\begin{figure*}
 \includegraphics[width=0.95\columnwidth]{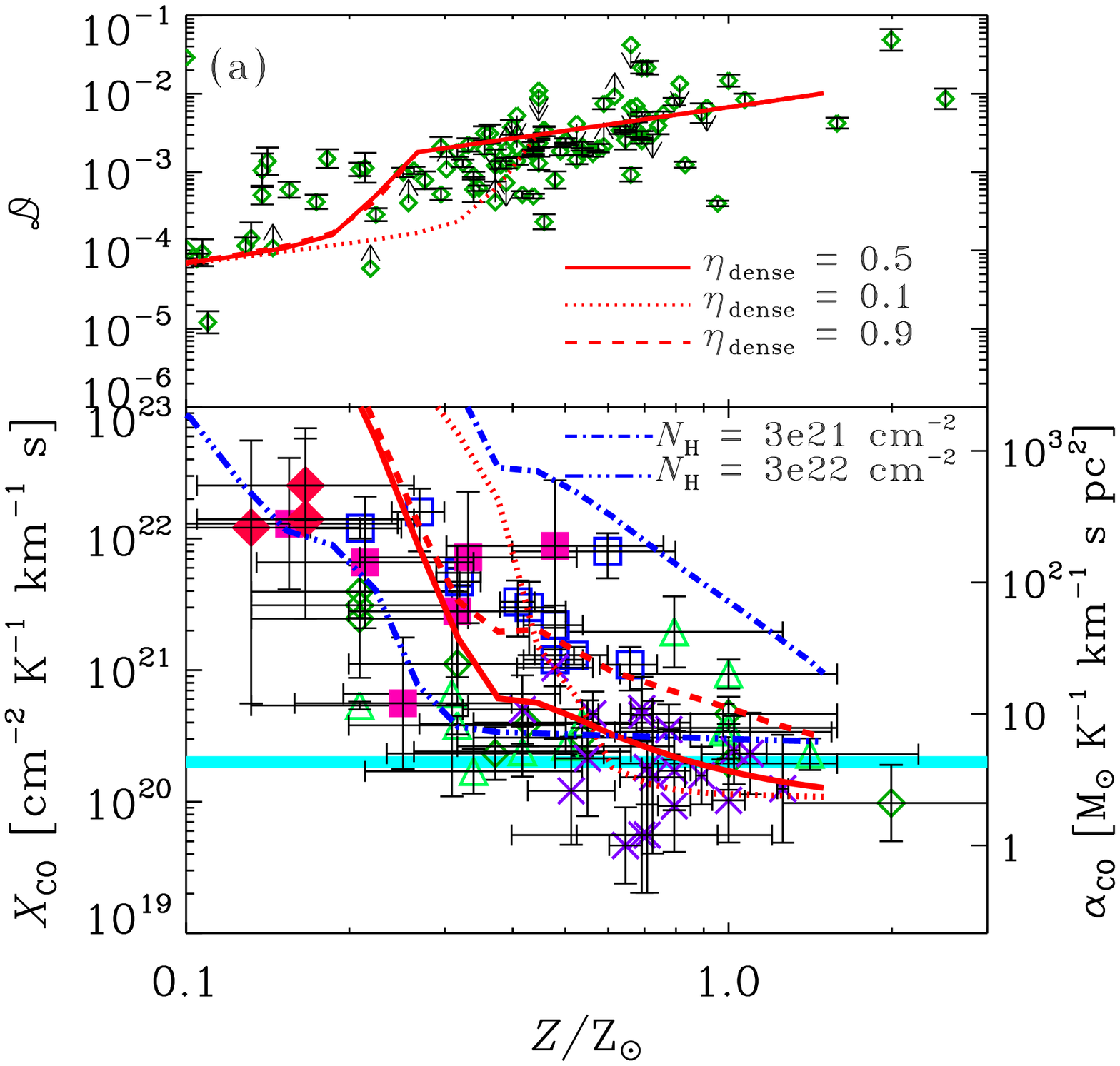}
 \includegraphics[width=0.95\columnwidth]{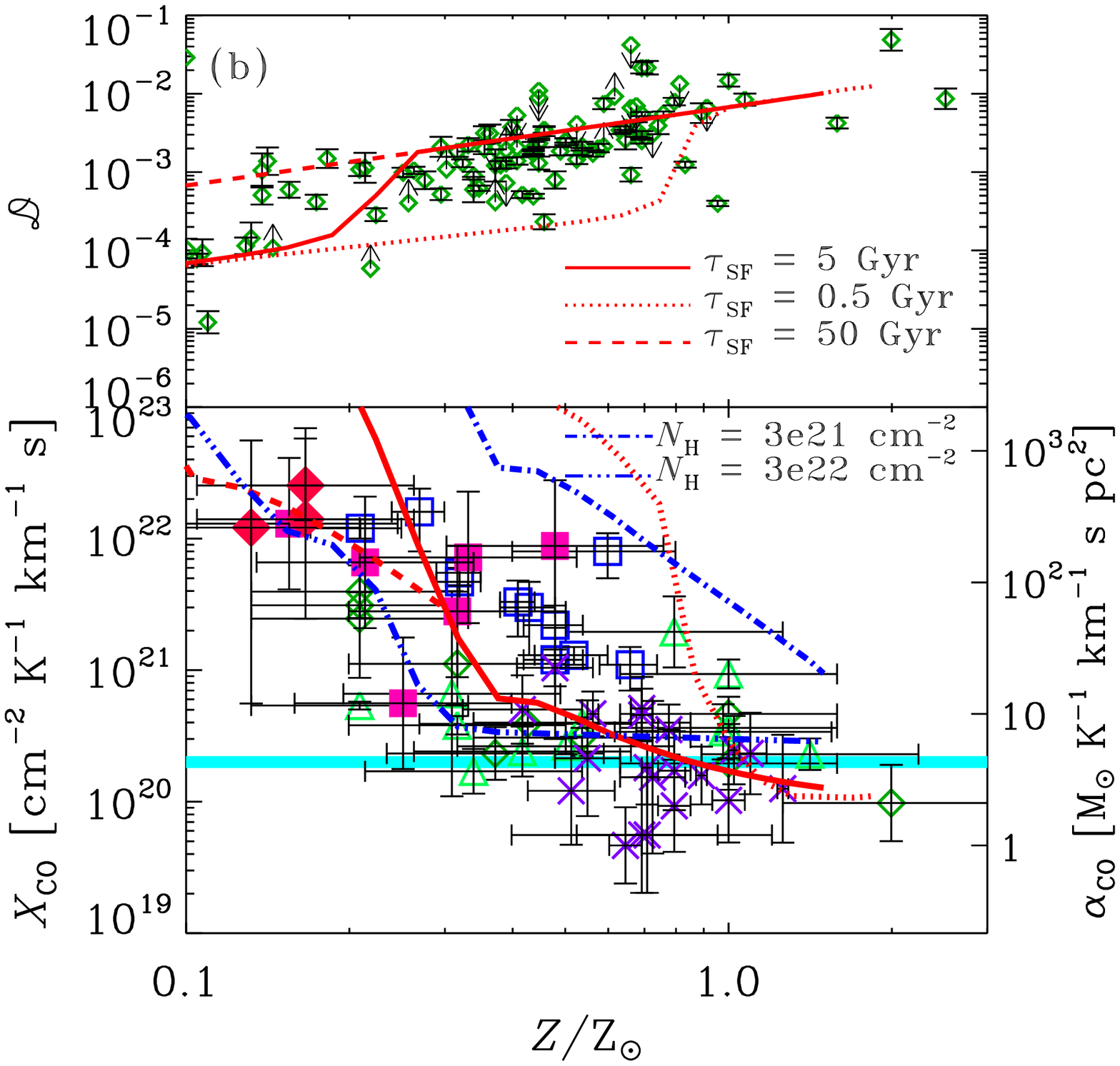}
 \caption{Relations between dust-to-gas ratio and metallicity (upper window in each panel)
 and between CO-to-H$_2$ conversion factor and metallicity (lower window).
 Panels (a) and (b) show the dependence on $\eta_\mathrm{dense}$ and
 $\tau_\mathrm{SF}$, whose values are the same as shown in Figs.\ \ref{fig:mol_eta} and \ref{fig:mol_tau},
 respectively (also shown in the legends).
 For $X_\mathrm{CO}$, we also plot the cases with lower and higher hydrogen column densities,
 $N_\mathrm{H}=3\times 10^{21}$ and $3\times 10^{22}$ cm$^{-2}$ by the dot--dashed and
 triple-dot--dashed lines, respectively,
 with $\eta_\mathrm{dense}=0.5$ and $\tau_\mathrm{SF}=5$ Gyr.
 The conversion factor is expressed in two units as shown on the left and
 right vertical axes.
 The points with error bars are the observation data in the literature:
 \citet{Remy-Ruyer:2014aa} in the upper window, and
 \citet[open diamonds]{Leroy:2011aa}, \citet[][triangles]{Bolatto:2008aa},
 \citet[open squares]{Israel:1997aa}, \citet[crosses]{Sandstrom:2013aa},
 \citet[filled squares]{Cormier:2014aa}, and \citet[filled diamonds]{Shi:2016aa}
 in the lower window, where we also show the MW value of $X_\mathrm{CO}$ by the
 horizontal thick line as a reference.}
 \label{fig:xco}
\end{figure*}

In Fig.\ \ref{fig:xco}, we compare our results with
the observational data for the $\mathcal{D}$--$Z$ and $X_\mathrm{CO}$--$Z$ relations.
We observe that
overall, the models nicely reproduce the increasing and decreasing trends
in $\mathcal{D}$ and $X_\mathrm{CO}$, respectively.
{For the $\mathcal{D}$--$Z$ relation, the nonlinear trend (or the steep increase of
$\mathcal{D}$ at subsolar metallicity) is caused by accretion, which is consistent with
the observed trend of dust-to-metal ratio \citep{Remy-Ruyer:2014aa}.}
The observed $X_\mathrm{CO}$--$Z$ relation is also explained by the same models;
in particular, the fiducial case is located in the middle of the observational data.
The variation among the cases with $\eta_\mathrm{dense}=0.1$--0.9 also explains
the scatter in the observed $\mathcal{D}$--$Z$ and $X_\mathrm{CO}$--$Z$ relations.
As mentioned in Section \ref{subsec:eta}, in spite of the almost same evolutionary track
in the $\mathcal{D}$--$Z$
relations between $\eta_\mathrm{dense}=0.5$ and 0.9,
these two cases have significantly different values of $X_\mathrm{CO}$
(originating from different $f_\mathrm{CO}$) at high metallicity
because of different grain size distributions.
Therefore, even if the dust-to-gas ratio is similar, the
CO-to-H$_2$ conversion factor can be very different because of different grain size distributions.
The high and low
values of $\tau_\mathrm{SF}$ explain the variations at high and low metallicities, respectively.
This is because the metallicity at which the prominent increase of $\mathcal{D}$ occurs
is shifted towards high and low metallicities for short and long $\tau_\mathrm{SF}$, respectively.

As shown in \citetalias{Hirashita:2017aa}, the hydrogen column density $N_\mathrm{H}$
of the cloud can also produce a large variation in the $X_\mathrm{CO}$--$Z$ relation.
Thus, in Fig.~\ref{fig:xco}, we consider an order of magnitude variation centred
at the fiducial value; that is, we examine
$N_\mathrm{H}=3\times 10^{21}$ and $3\times 10^{22}$~cm$^{-2}$
in addition to the fiducial case.
Note that $N_\mathrm{H}$ does not affect the $\mathcal{D}$--$Z$ relation.
We overall find that the above range of $N_\mathrm{H}$ is consistent with the range (or the scatter)
of $X_\mathrm{CO}$ in the observational data.
As expected, the larger and smaller values of $N_\mathrm{H}$ predict higher and lower
values of $f_\mathrm{CO}$, leading to lower and higher values of $X_\mathrm{CO}$,
respectively, except at high metallicity.
At solar metallicity and above, $X_\mathrm{CO}$ is higher for
$N_\mathrm{H}=3\times 10^{22}$ cm$^{-2}$ than for $N_\mathrm{H}=10^{22}$ cm$^{-2}$
because the saturation of the CO line intensity due to high optical depth
is more prominent at higher $N_\mathrm{H}$. This means that, at high metallicity,
$N_\mathrm{H}\sim 10^{22}$ cm$^{-2}$ is the optimum column density for the CO $J=1\to 0$
emission intensity per hydrogen, and the higher and lower column densities both
lead to less efficient emission.

In Fig.\ \ref{fig:xco}, we also show the MW value of $X_\mathrm{CO}$ as a reference.
We observe that the fiducial model ($\eta_\mathrm{dense}=0.5$ and $\tau_\mathrm{SF}=5$ Gyr)
reaches the MW value at solar metallicity, which is appropriate for the metallicity of the MW.
This confirms that our grain evolution model predicts a conversion factor consistent with the
observed one at solar metallicity if we choose standard values for
relevant parameters.
%%($n_\mathrm{H}$, $N_\mathrm{H}$, $\chi$, $\Delta v$, and $T_\mathrm{cl}$).
{We should still be aware of possible systematic errors that could cause
offsets for the observationally obtained $X_\mathrm{CO}$. At the same time,
our model also contains adjustable parameters such as $T_\mathrm{cl}$ and
$\Delta v$. In spite of these systematic errors and adjustments, we still firmly conclude that
our model is capable of reproducing the trend in the $X_\mathrm{CO}$--$Z$ relation
well.}

As expected from the above results, a large fraction
of carbon atoms in dense gas are traced by emission not from CO but from
other forms of carbon
such as C \textsc{ii} and C \textsc{i} at low metallicity
\citep[e.g.][]{Madden:1997aa,Cormier:2014aa,Glover:2016aa,Hu:2021aa}.
The transition of C \textsc{ii}/C \textsc{i}-dominated to
CO-dominated carbon content occurs at $Z\sim 0.2$--0.3 Z$_{\sun}$ in our fiducial model.
This corresponds to the metallicity at which dust growth
by accretion causes a rapid increase of the dust abundance.
We note that dust shielding, which helps CO formation,
is further enhanced by efficient small-grain production
around this metallicity.

\section{Discussion}\label{sec:discussion}

\subsection{Dust-based conversion factor formulae}

The above results indicate that the dust abundance and the grain size distribution
have large impact on the CO-to-H$_2$ conversion factor.
This motivates us to further analyze the relation between $X_\mathrm{CO}$ and
dust-related quantities. Here we reanalyze our results in terms of the dust-to-gas ratio
$\mathcal{D}$ and the grain size distribution.

\begin{figure*}
 \includegraphics[width=0.95\columnwidth]{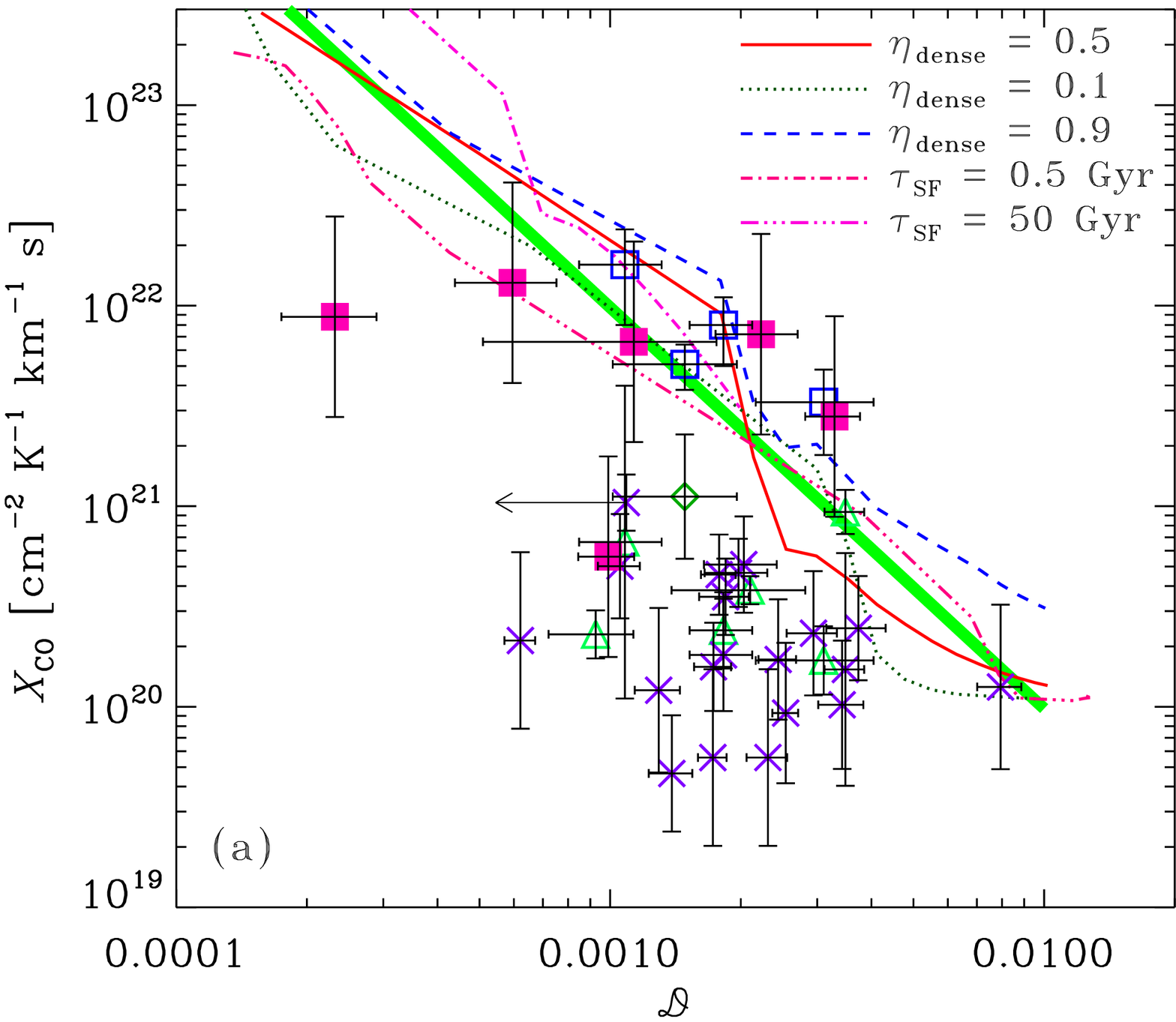}
 \includegraphics[width=0.95\columnwidth]{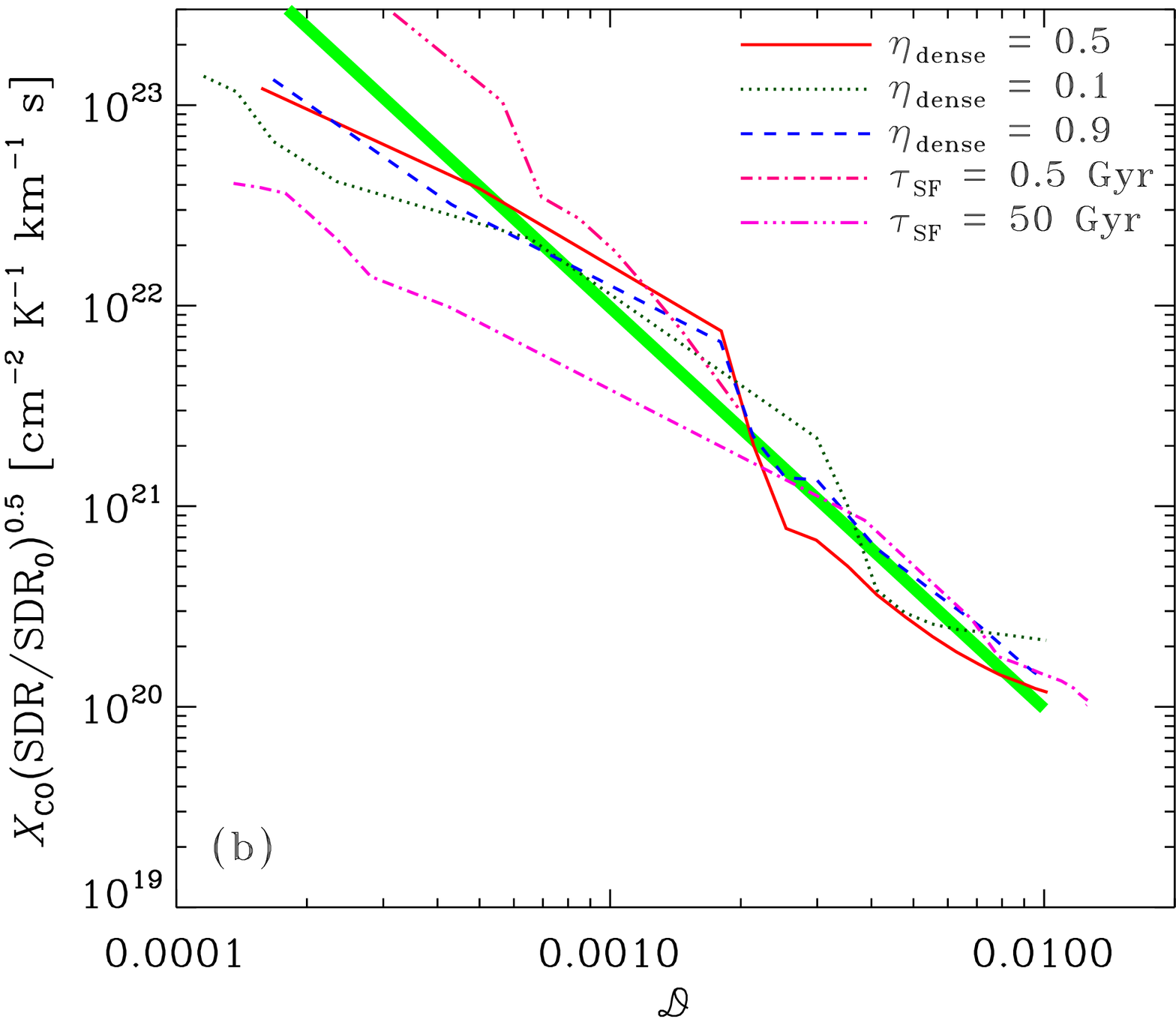}
 \caption{(a) CO-to-H$_2$ conversion factor as a function of dust-to-gas ratio.
 The (thin) solid, dotted, and dashed curves present the cases with $\eta_\mathrm{dense}=0.5$
 (fiducial), 0.1, and 0.9, respectively, while the dot--dashed, and triple-dot--dashed lines
 show the results with $\tau_\mathrm{SF}=0.5$ and 5 Gyr, respectively.
 The parameters other than the varied one are fixed to the fiducial values.
 The thick light green
 line is the proposed power-law formula in equation (\ref{eq:Xco_dg}).
 (b) The relations shown is Panel (a) is corrected for
 the $(\mathrm{SDR}/\mathrm{SDR}_0)^{0.5}$ factor, which includes the information
 on the grain size distribution, to minimize the variety at high metallicity. The models are the same as
 in Panel (a), and the thick green line shows the proposed power-law formula
 in equation (\ref{eq:Xco_dg_corr}) with $\mathrm{SDR}=\mathrm{SDR}_0$.}
 \label{fig:co_dg}
\end{figure*}

First we examine how $X_\mathrm{CO}$ is related to the dust abundance represented by
the dust-to-gas ratio $\mathcal{D}$.
In Fig.~\ref{fig:co_dg}a, we show $X_\mathrm{CO}$
as a function of dust-to-gas ratio $\mathcal{D}$.
We present various models investigated above:
the variation of $\eta_\mathrm{dense}$ and $\tau_\mathrm{SF}$.
The parameters other than the varied one are fixed to the fiducial values.
%%$f_\mathrm{CO}$ still has a large dispersion at high metallicity,
%%reflecting the variety in the grain size distribution.
%%This confirms, as shown above, that the CO abundance is sensitive to the grain size distribution
%%even if the dust-to-gas ratio is the same.
We observe in the figure that the conversion factor $X_\mathrm{CO}$ is better aligned in
a single sequence if we plot it as a function of $\mathcal{D}$ instead of $Z$.
Overall, $X_\mathrm{CO}$ can be approximated by a power-law function of $\mathcal{D}$
with a single slope, while it had a clear kink when it was plotted as a function of $Z$
(Figs.~\ref{fig:mol_eta} and \ref{fig:mol_tau}).
We suggest the following formula roughly reproducing the trends in all the models shown:
\begin{align}
X_\mathrm{CO}=2\times 10^{20}\left(\frac{\mathcal{D}}{7\times 10^{-3}}\right)^{-2}~
\mathrm{cm^{-2}~\mathrm{K}^{-1}~km^{-1}~s},\label{eq:Xco_dg}
\end{align}
which is obtained with a constraint that the MW dust-to-gas ratio $\mathcal{D}=7\times 10^{-3}$
\citep{Weingartner:2001aa,Hirashita:2023aa} reproduces the MW value of $X_\mathrm{CO}$
as well as a requirement that the overall trend traces all the models shown.
The conversion factors in all the models are broadly
consistent with this suggested formula
%%(equation~\ref{eq:Xco_dg})
within a factor of 3.

We also show the observational data for comparison in Fig.\ \ref{fig:co_dg}a.
We adopt the same sample as in Fig.\ \ref{fig:xco} but only plot galaxies whose dust-to-gas ratio
is available from \citet{Remy-Ruyer:2014aa}; i.e.\ we use the consistent values of dust-to-gas ratio
with the upper windows of Fig.\ \ref{fig:xco}.
We note that the comparison is not fully consistent because
most observational data are analyzed with different assumptions on the dust-to-metal
(or dust-to-gas) ratio in evaluating $X_\mathrm{CO}$ or
on the CO-to-H$_2$ conversion factor in estimating $\mathcal{D}$.
For uniformity of the data, the estimated values of $\mathcal{D}$ are taken from
a single paper.
We observe that the data from \citet{Sandstrom:2013aa}, which are shown by crosses,
are systematically located at
lower $X_\mathrm{CO}$ compared with the points from
\citet{Israel:1997aa} and \citet{Cormier:2014aa}.
Note that the latter two papers assumed relations between dust and gas masses (or emissions).
%%Although these assumptions may affect the comparison, we leave it for future work.
\citet{Sandstrom:2013aa} did not assume any relation between gas and dust masses,
and find self-consistent values of $X_\mathrm{CO}$ and $\mathcal{D}$ in their
spatially resolved galaxy maps under an assumption
that the correct solution should minimize the dispersion of $\mathcal{D}$ within a kpc-scale region.
Although this assumption is plausible, a further study is necessary to understand
possible bias by cross-checking various methods.

We showed above that the grain size distribution plays an important role in
the CO abundance. To obtain a simple quantity that characterizes
the grain size distribution,
we first define the following weighted integral of a function $f(a)$ for
the grain size distribution:
\begin{align}
I(f)\equiv\sum_i\int_0^\infty f_i(a)n_i(a)\,\mathrm{d}a,
\end{align}
where the index $i$ specifies the grain species.
%%We aim at extracting the quantity that characterizes the grain size distribution most
%%efficiently.
A simple example is a series of moments: $f(a)=a^\ell$ ($\ell =0$, 1, 2, $\cdots$)
as used by \citet{Mattsson:2016aa}. Although the full moment treatment is out of the scope
of this paper, a simple physical intuition may lead to the importance of
the second and third moments ($\ell =2$ and 3), which are indicators of the
grain surface area and volume, respectively. For the purpose of this paper,
we take $S_\mathrm{d}(a)=\upi a^2$ and $m_\mathrm{d}(a)=4\pi a^3s/3$
for the function $f(a)$ above. The dust-to-gas ratio is expressed as
$\mathcal{D}=I(m_\mathrm{d})/(1.4m_\mathrm{H}n_\mathrm{H})$ and the grain
surface area per gas mass, denoted as $\mathcal{S}$, is calculated by
$\mathcal{S}=I(S_\mathrm{d})/(1.4m_\mathrm{H}n_\mathrm{H})$, where
the factor 1.4 indicates the correction for helium. Using these two quantities,
the surface-to-mass ratio of the dust is obtained as $\mathcal{S}/\mathcal{D}$.
Because the grain surface area can be used as an indicator of the opacity for dissociating radiation
(although the relation is not completely proportional), we try to use the `SD ratio' defined as
$\mathrm{SDR}\equiv\mathcal{S}/\mathcal{D}$
as a quantity that reflects the main effect of the grain size distribution.

Since $X_\mathrm{CO}$ is affected by the grain size distribution in addition to the dust abundance,
we propose that $X_\mathrm{CO}\propto (\mathrm{SDR})^\gamma F(\mathcal{D})$,
where $F(\mathcal{D})$ is a function of $\mathcal{D}$.
In other words, the explicit dependence on the grain size distribution is assumed to be
expressed by a power-law function of SDR for simplicity. With this assumption,
$X_\mathrm{CO}(\mathrm{SDR})^{-\gamma}$ is a function of $\mathcal{D}$
not explicitly dependent on the shape of grain size distribution.
After some tests, $\gamma\sim -0.5$ minimizes the
dispersion among the models at high metallicity (where the CO detection is
actually expected).
%%In Fig.\ \ref{fig:co_dg}b, we show $X_\mathrm{CO}(\mathrm{SDR})^{-0.5}$
%%as a function of $\mathcal{D}$.
In our fiducial model, $\mathrm{SDR}=2.9\times 10^5$ cm$^{2}$ g$^{-1}$
at the MW dust-to-gas ratio ($\mathcal{D}=7\times 10^{-3}$).

In Fig.~\ref{fig:co_dg}, we show $X_\mathrm{CO}(\mathrm{SDR}/\mathrm{SDR}_0)^{0.5}$,
where $\mathrm{SDR}_0=2.9\times 10^5$ cm$^{2}$ g$^{-1}$ is used for normalization
based on our fiducial model, as a function of $\mathcal{D}$.
We confirm that the dispersion among the models at high metallicity becomes
significantly smaller if we present
$X_\mathrm{CO}(\mathrm{SDR}/\mathrm{SDR}_0)^{0.5}$ instead of $X_\mathrm{CO}$.
In the figure, we also show the corrected version of equation (\ref{eq:Xco_dg}):
\begin{align}
X_\mathrm{CO}=2\times 10^{20}\left(\frac{\mathcal{D}}{7\times 10^{-3}}\right)^{-2}~
\left(\frac{\mathrm{SDR}}{\mathrm{SDR}_0}\right)^{-0.5}
\mathrm{cm^{-2}~\mathrm{K}^{-1}~km^{-1}~s}.\label{eq:Xco_dg_corr}
\end{align}
This formula is useful if both dust-to-gas ratio
and SDR are available or can be reasonably assumed.
It is, however, usually difficult to obtain the information on the dust surface area;
in this case, we could fix SDR to our fiducial value ($\mathrm{SDR}_0$), and
the relation is reduced to equation (\ref{eq:Xco_dg}).

Since dust plays a more prominent role in regulating the CO abundance than metals,
it may be more robust to use the dust-based formulae proposed in
equation (\ref{eq:Xco_dg}) or (\ref{eq:Xco_dg_corr}).
However, the usefulness of these formulae is not obvious in the following two
points: (i) For the gas mass estimate necessary to obtain the dust-to-gas ratio,
we need $X_\mathrm{CO}$; thus, the process of estimating dust-to-gas ratio
with an undetermined value of $X_\mathrm{CO}$ is iterative. (ii) The relation is not yet
observationally supported in a robust way.
Seeing Fig.\ \ref{fig:co_dg}a, there are a significant number
of data points deviating from the proposed relation. However, the observational estimate of
$X_\mathrm{CO}$ also needs some assumptions.
Thus, further cross-checks among the observational methods of deriving
$X_\mathrm{CO}$ are necessary to understand possible
systematic errors in each data set.

\subsection{Possible dependence on other parameters}\label{subsec:others}

As mentioned in Section~\ref{subsec:choice}, some of the physical parameters for
the typical cloud may affect the results. In particular, the gas temperature and
the ISRF are expected to have a large variety depending on the star formation
activity and the detailed spatial distributions of stars and clouds in the galaxy.
Thus, we examine the variation of cloud parameters.
Since we already investigated the effect of $N_\mathrm{H}$ in
Section \ref{subsec:comparison},
we fix it to the fiducial value ($N_\mathrm{H}=10^{22}$ cm$^{-2}$).

\begin{figure}
 \includegraphics[width=1\columnwidth]{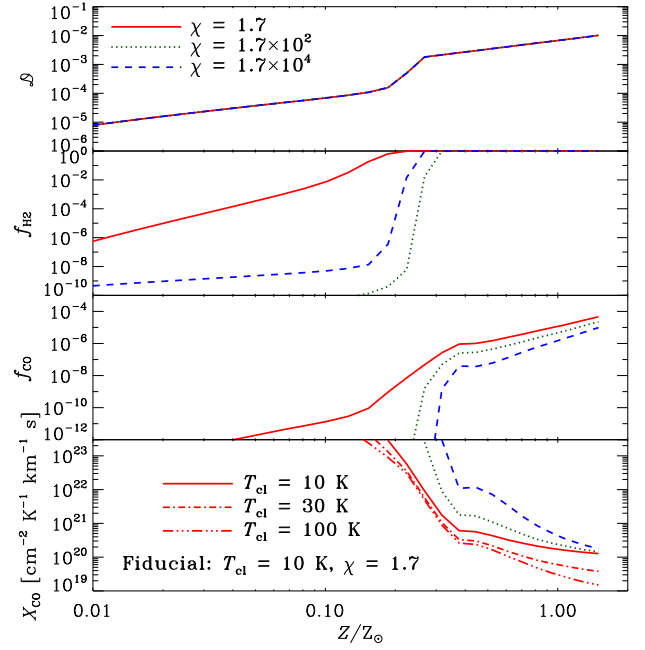}
 \caption{Same as Fig.\ \ref{fig:mol_eta} but for different values of
 for $\chi$ (1.7, $1.7\times 10^2$ and $1.7\times 10^4$; solid, dotted and
 dashed lines, respectively) with $T_\mathrm{cl}=10$ K.
 We also show $X_\mathrm{CO}$ for $T_\mathrm{cl}=30$ and 100 K with
 $\chi =1.7$ (dot--dashed and triple-dot--dashed lines, respectively).
 Note that $T_\mathrm{cl}$ prominently affects $X_\mathrm{CO}$, so
 that the $T_\mathrm{cl}$ dependence is only shown in the bottom panel.
 Neither $\chi$ nor $T_\mathrm{cl}$ influences the
 evolution of $\mathcal{D}$. For the parameters other than the varied one, we
 adopt the fiducial values.}
 \label{fig:highchi}
\end{figure}

Starburst galaxies have SFR surface densities up to
$\sim 10^4$ times higher than normal spiral galaxies
\citep{Kennicutt:1998aa}.
Assuming that the ISRF intensity is proportional to the SFR surface density,
we examine $\chi$ up to $1.7\times 10^4$. In Fig.\ \ref{fig:highchi},
we show the results for $\chi =1.7$, $1.7\times 10^2$ and $1.7\times 10^4$
with other parameters fixed to the fiducial values.
Note that the evolution of $\mathcal{D}$ is not affected by $\chi$.
We observe in the figure that the H$_2$ and CO abundances are suppressed
in high radiation fields at low metallicity.
The cloud, however, achieves $f_\mathrm{H_2}\sim 1$ at
$Z\gtrsim 0.3$ Z$_{\sun}$, and this metallicity is not sensitive to $\chi$.
The ISRF intensity also affects $f_\mathrm{CO}$ at low metallicity,
but its effect is less at high metallicity because a large fraction of the ISRF is
shielded by dust.
Accordingly, the CO-to-H$_2$ conversion factor has a large variety at low
metallicity, while the difference becomes moderate at high metallicity.
Indeed, $X_\mathrm{CO}$ varies only by a factor of $\sim$3
at solar metallicity although $\chi$ differs by four orders of magnitude.

In Fig.\ \ref{fig:highchi}, we also show the dependence on $T_\mathrm{cl}$.
Note that the prominent effect of $T_\mathrm{cl}$ is only seen for
the CO-to-H$_2$ conversion factor in our model, so that the $T_\mathrm{cl}$-dependence is only
shown for $X_\mathrm{CO}$.
We observe that $T_\mathrm{cl}$ little affects $X_\mathrm{CO}$ at low
metallicity because $W_\mathrm{CO}$ is insensitive to $T_\mathrm{cl}$
(Section \ref{subsec:choice}).
At high metallicity, the line emission becomes
optically thick, which leads to the CO line intensity almost proportional to
$T_\mathrm{cl}$. Thus, $X_\mathrm{CO}$ is lower for higher $T_\mathrm{cl}$
at high metallicity.
Variation of $\Delta v$ (not shown in the figure) also has a similar effect to that of $T_\mathrm{cl}$.

Starburst galaxies such as (ultra)luminous infrared galaxies have lower
$X_\mathrm{CO}\sim 0.2$--$1\times 10^{20}$ cm$^{-2}$ K$^{-1}$ km$^{-1}$ s,
than that in the MW \citep[see section 7 of][for a review]{Bolatto:2013aa}.
As supported by the above results, the low conversion factors in starburst galaxies could be
explained by high gas temperatures \citep{Wild:1992aa} or high velocity dispersions
\citep{Zhu:2003aa,Papadopoulos:2012aa}
or both \citep{Narayanan:2011aa}. The high radiation field may have a less significant
influence on $X_\mathrm{CO}$ if these galaxies have nearly
solar metallicity.

Although our model has been compared with nearby galaxies,
it is also applicable to high-redshift galaxies.
\citet{Tacconi:2008aa} observed CO lines from submillimetre galaxies (SMGs)
and UV/optically selected galaxies at $z\sim 2$. Their most favoured solution
indicates that the CO-to-H$_2$ conversion factors of the SMGs are similar to
nearby ULIRGs while those of the UV/optically selected galaxies are near to the
MW value.
\citet{Daddi:2010aa} also showed that the CO-to-H$_2$ conversion factors of
the main-sequence galaxies
at $z\sim 1.5$ are near the MW value.
\citet{Magdis:2011aa} also found
a difference in the conversion factor between a SMG and
a main-sequence galaxy.
\citet{Magnelli:2012aa} showed a negative correlation between
CO-to-H$_2$ conversion factor and dust temperature for galaxies
at $z\sim 1$. The dust temperature is expected to be positively
correlated with $\chi$; however, higher $\chi$ predicts larger $X_\mathrm{CO}$
in our model (Fig.\ \ref{fig:highchi}), which is opposite to the observed trend.
Since their sample has solar metallicity,
the difference in $X_\mathrm{CO}$ could be more prominently caused by
the variation in gas temperature or velocity dispersion as discussed above
\citep[see also][]{Maloney:1988aa}.
%%The dust temperature is also linked to the dust-to-gas ratio, since
%%high dust abundance leads to lower dust temperature because of more dust shielding
%%\citep{Liang:2019aa,Sommovigo:2022aa,Hirashita:2022ab}.
%%Therefore, the observational correlation between dust temperature and $X_\mathrm{CO}$
%%could also be caused by the negative relation between $\mathcal{D}$ and $X_\mathrm{CO}$.

Our results also imply difficulty in detecting CO at $z\gtrsim 5$ when the
cosmic age is $\lesssim 1$ Gyr.
We expect that many of the galaxies at $z\gtrsim 5$ are metal-poor, which
indicates that CO is difficult to detect because $X_\mathrm{CO}$ is large.
Our result
in Section \ref{subsec:tau_SF} further indicates that galaxies with short
$\tau_\mathrm{SF}$ have a sharp transition from large to small $X_\mathrm{CO}$ at
solar metallicity. Thus, even if a galaxy
at $z\gtrsim 5$ experiences a quick metal enrichment (i.e.\
has a short $\tau_\mathrm{SF}$), it is not necessarily
CO-rich. The transition from a CO-poor to CO-rich galaxy occurs in
a narrow range of metallicity. This may predict that high-redshift galaxies
are `bimodal' in the CO-rich and CO-poor phases.
Since this transition is related to the dust growth by accretion,
we also predict that this CO-rich/poor bimodality is associated with the
dust-rich/poor distinction.
If the fraction of galaxies with short $\tau_\mathrm{SF}$ is large at high redshift,
it is important to use [C \textsc{ii}] and/or [C \textsc{i}] emission to
trace star-forming gas in galaxies.

\subsection{Conversion factor prescriptions}\label{subsec:prescription}

Most of the prescriptions for the conversion factor include the metallicity dependence.
Many of them
adopted a power-law metallicity dependence of $X_\mathrm{CO}$
\citep[e.g.][]{Israel:1997aa,Schruba:2012aa,Hunt:2015aa,Accurso:2017aa}.
These power-law dependences are broadly consistent with the observational data
shown in Fig.\ \ref{fig:xco}.
From a physical insight from the fraction of CO-dark molecular gas, exponential 
dependence may be expected \citep{Wolfire:2010aa,Bolatto:2013aa}.
Because of the large scatter in the observational data, we are not able to
judge which functional form of $X_\mathrm{CO}$ describes the observations
better.

Our results show that the $X_\mathrm{CO}$--$Z$ relation is also affected
by the
dust evolution.
If we focus on $Z\gtrsim 0.3$ Z$_{\sun}$, $X_\mathrm{CO}$ could be approximated
by a power law, while the sharp rise towards lower metallicity may be better described by
a steeper function.
%%In particular, as shown in Fig.\ \ref{fig:xco},
%%the $X_\mathrm{CO}$--$Z$ relation has a sharp drop at the metallicity where
%%dust growth by accretion start to become the dominant dust source.
%%After this, the relation has a kink and transit to a milder drop.
The sharp decrease of $X_\mathrm{CO}$ is associated with the change of
the major dust sources from stellar dust production to dust growth by accretion.
Thus, to describe the $X_\mathrm{CO}$--$Z$ relation in a wide metallicity range,
it is crucial to understand or model the dust evolution.

Note again that $X_\mathrm{CO}$ also depends on quantities other than
the metallicity and the grain size distribution.
\citet{Bolatto:2013aa} suggested that $X_\mathrm{CO}$ depends on the
surface density of total baryons ($\Sigma_\mathrm{tot}$)
as $X_\mathrm{CO}\propto\Sigma_\mathrm{tot}^{-0.5}$ at high
$\Sigma_\mathrm{tot}(>100~\mathrm{M}_{\sun}~\mathrm{pc}^{-2})$
such as realized in galaxy centres.
\citet{Chiang:2021aa} showed that the $\Sigma_\mathrm{tot}$-dependent
$X_\mathrm{CO}$ is preferred to obtain reasonable radial profiles of dust-to-metal
ratio for a sample of nearby galaxies.
The $\Sigma_\mathrm{tot}$ dependence is
probably due to the difference in the physical conditions regulated by the gravity.
In particular, as shown above, the temperature and the velocity dispersion
of the cloud affect $X_\mathrm{CO}$ if CO is optically thick
(Section~\ref{subsec:others}).
It is natural to consider that $\Delta v^2$ and $T_\mathrm{cl}$
reflect the depth of the gravitational potential,
which is proportional to $\Sigma_\mathrm{tot}$.
Thus, we hypothesize that $\Delta v^2\propto \Sigma_\mathrm{tot}^p$ and
$T_\mathrm{cl}\propto\Sigma_\mathrm{tot}^q$ with $p>0$ and $q>0$.
We also assume that the molecular gas surface density, $\Sigma_\mathrm{mol}$
positively correlates with $\Sigma_\mathrm{tot}$: $\Sigma_\mathrm{mol}\propto\Sigma_\mathrm{tot}^r$
with $r>0$. With these assumptions, we obtain
$X_\mathrm{CO}\propto\Sigma_\mathrm{mol}/(T_\mathrm{cl}\Delta v)\propto\Sigma_\mathrm{tot}^{(r-0.5p-q)}$
at high optical depth (at high metallicity). 
Thus, if $r-0.5p-q=-0.5$, the above surface density dependence can be explained.
For example, if we assume simple proportionality for all the
relations (i.e.\ $p=q=r=1$), the above relation is satisfied.
Although further sophisticated dynamical modelling would be required to obtain
more precise values of $p$, $q$ and $r$
\citep[see also section 2 of][]{Bolatto:2013aa}, the above simple argument implies
that the dependence on the total surface density can be translated
into that on the gas density and velocity dispersion investigated in Section \ref{subsec:others}.

Our one-zone approach is not capable of calculating the physical conditions of
dense clouds in a consistent manner with the hydrodynamic evolution of the ISM.
Hydrodynamic simulations provide a viable method for predicting physical quantities
that govern the H$_2$ and CO abundances. Some hydrodynamic simulations of galaxies included
the evolution of grain size distribution. In particular, \citet{Aoyama:2020aa} showed that
the grain size distributions are systematically different between the dense and diffuse ISM.
Thus, our assumption of identical grain size distribution everywhere in the galaxy may need
to be modified. However, \citet{Romano:2022ab}, using the same
simulation framework but including turbulent diffusion, showed that the grain size distribution can
be homogenized between the dense and diffuse ISM. This means that a one-zone model could
provide a reasonable description if the diffusion is strong.
The simple one-zone study in this paper will give a basis
on which we interpret spatially resolved evolution in future simulations.

{Another feature that a one-zone treatment is not able to predict is the variation
within a galaxy. \citet{Hou:2017aa} showed that the small-to-large grain abundance ratio
varies with galactocentric distance \citep[see e.g.][for a recent simulation]{Romano:2022ab}.
This variation is mainly driven by different metal enrichment
(or star formation) histories. However, the evolutionary sequence of grain size distribution is
similar to that predicted from a one-zone model with a delay in regions with slower
metal enrichment (usually in the outer galactic discs).
The spatial variation of CO-to-H$_2$ conversion factor
is also an important topic to clarify using a frameworks that combines
our models developed in this paper and hydrodynamic simulations.}

\subsection{Possible impacts of inhomogeneity in the cloud}\label{subsec:inhomogeneity}

{Our treatment of H$_2$ abundance is based on a uniform density and
a homogeneous mixture between dust and gas. There are some effects that inhomogeneity
could have on the H$_2$ abundance. Broadly, there are two types of effects: One
is the effect of local density enhancement and the other is dust--gas decoupling.}

{In an inhomogeneous medium, the reaction rate is enhanced in regions
where the density is higher than the average. The H$_2$ formation, which is
proportional to the product of dust and gas densities, is enhanced by a factor
of $\langle n_\mathrm{H}n_\mathrm{d}\rangle /\langle n_\mathrm{d}\rangle\langle
n_\mathrm{H}\rangle$, where the bracket
indicates the spatial average within the cloud, and $n_\mathrm{H}$ and $n_\mathrm{d}$ are
the local number densities of hydrogen nuclei and dust grains, respectively.
If dust and gas are tightly coupled (i.e.\ the dust-to-gas ratio is uniform),
the above ratio is reduced to
$\langle n_\mathrm{H}^2\rangle /\langle n_\mathrm{H}\rangle^2$.
The density enhancement is related to the mean Mach number
of the turbulence \citep[e.g.][]{Vazquez:1994aa,Federrath:2008aa};
that is, the density enhancement is a dynamical phenomenon. Thus, we need to
take into account the finite formation time of H$_2$, and evaluate the H$_2$ fraction
in a manner consistent with the dynamical density evolution within the cloud.
Although this dynamical treatment is not possible in our framework, our formulae for
the H$_2$ formation rate including the effect of grain size distribution is generally
applicable. Future development that combines the grain size distribution and
the hydrodynamic evolution is necessary to address the local enhancement of
H$_2$ formation rate in an inhomogeneous structure.}

{The assumption of homogeneous mixing between dust and gas also needs to be
checked carefully. \citet{Hopkins:2016aa} showed that decoupling of dust grains from
small-scale gas density structures could make significantly different spatial distributions
between dust and gas. The resulting spatial distribution of dust is, however,
affected by complex factors such as
grain charging and magnetic field \citep{Lee:2017aa,Beitia:2021aa,Moseley:2023aa}.
Basically, the effect of decoupling would weaken the effect of local density enhancement:
The enhancement factor for the H$_2$ formation rate is written as
$\langle n_\mathrm{H}n_\mathrm{d}\rangle /\langle n_\mathrm{d}\rangle\langle
n_\mathrm{H}\rangle$ as mentioned above, and is reduced to unity if dust and gas are independently
distributed (i.e.\ $\langle n_\mathrm{H}n_\mathrm{d}\rangle =\langle n_\mathrm{H}\rangle\langle
n_\mathrm{d}\rangle$). Thus, decoupling would tend to justify the usage of
averaged quantities. However, detailed effects depend on the
grain radius, so that it is interesting to further investigate the effect of decoupling on the
H$_2$ formation by combining our grain size distribution model
and dust--gas dynamic simulations in future work.
}

{Compared with the formation rate, the dissociation rate of H$_2$ is less
affected by inhomogeneity in
density and dust-to-gas ratio for the following reasons.
The optical
depth for dissociating radiation reflects the mean density along the light path.
although the precise shielding strength
depends on the clumpiness and the optical depth of each clump
\citep[e.g.][]{Varosi:1999aa}.
The formation rate, in contrast, depends directly on the local density.
Thus, we expect that density inhomogeneity affects the formation rate of H$_2$
much more than the destruction rate.
Moreover, as mentioned in
Section \ref{subsec:eta}, self-shielding is the main shielding mechanism for H$_2$.
Therefore, the detailed spatial distribution of dust
is not important for shielding, which means that
dust--gas decoupling has a minor influence on H$_2$ dissociation.}

{From the above discussions, it is possible that density inhomogeneity
could enhance the H$_2$ abundance through the local enhancement of
H$_2$ formation rate. Therefore, $f_\mathrm{H_2}$ could reach unity at
lower metallicity than our results. This does not change our
conclusion for $X_\mathrm{CO}$ since $f_\mathrm{H_2}$ reaches unity in the
metallicity range of interest ($Z\gtrsim 0.1$ Z$_{\sun}$) even in
our treatment with homogeneous density.}

{For the CO abundance, since we adopted the results of hydrodynamic simulations
from \citetalias{Glover:2011aa},
we effectively included density inhomogeneity as mentioned in
Section \ref{subsec:fco}. We still formulated dust shielding of
CO in a one-zone treatment; as argued above, however, shielding is not
strongly affected by density inhomogeneity. After all, it is not likely that
our results for $X_\mathrm{CO}$
are largely affected by the assumption of homogeneity adopted in this paper.}

\section{Conclusions}\label{sec:conclusion}

We investigate how the evolution of grain size distribution affects the
H$_2$ and CO abundances. Our model is based on
\citetalias{Hirashita:2017aa}, but including the full treatment of
grain size distribution.
The calculation of grain size distribution is performed using the
\citetalias{Hirashita:2020aa} model but is modified to treat silicate and carbonaceous dust
separately.
This model includes the following processes: stellar dust production,
dust destruction in supernova shocks, dust growth by accretion and coagulation,
and grain disruption by shattering.
We treat the galaxy as a one-zone object,
assuming that the grain size distribution is the same
in any part of the galaxy.
The H$_2$ formation rate and
the shielding (extinction) efficiency of dissociating radiation for H$_2$ and CO
are evaluated in a manner
consistent with the calculated grain size distribution at each epoch.
To concentrate on the effect of grain size distribution,
we basically fix the physical condition of a typical cloud as 
$N_\mathrm{H}=10^{22}$~cm$^{-2}$, $n_\mathrm{H,cl}=10^3$ cm$^{-3}$,
and $T_\mathrm{cl}=10$ K.
We show the evolution of the H$_2$ and CO abundances
and the CO-to-H$_2$ conversion factor $X_\mathrm{CO}$ as a function of metallicity.

We find that the H$_2$ fraction ($f_\mathrm{H_2}$) increases drastically in the
epoch when the total grain surface area increases owing to small grain
production by shattering.
The formed H$_2$ further self-shields the
dissociating radiation, accelerating the increase of H$_2$ abundance.
The cloud becomes fully molecular even before
dust growth by accretion significantly raises the dust abundance.
The increase of dust abundance is important for CO, whose abundance is strongly
regulated by dust shielding. Therefore, the increase of dust-to-gas ratio by accretion
drives the drop of $X_\mathrm{CO}$. After accretion is saturated, the dust-to-gas ratio
only linearly increases as a function of $Z$. At this epoch, the increase of $f_\mathrm{CO}$
and the decrease of
$X_\mathrm{CO}$ as a function of $Z$ becomes milder.
The metallicity dependence of $X_\mathrm{CO}$ is broadly consistent with the observational
data, but it is not described by a simple or single power-law.

The evolution of grain size distribution can be regulated by changing the dense gas
fraction $\eta_\mathrm{dense}$ in our model.
In particular, if $\eta_\mathrm{dense}$
is as small as 0.1, accretion becomes efficient at higher metallicity than in the
case of larger $\eta_\mathrm{dense}$ because the abundance of dense gas hosting accretion
is low. However, once the dust abundance is increased by accretion, the CO abundance is higher
for $\eta_\mathrm{dense}=0.1$ than for larger values of $\eta_\mathrm{dense}$ because
the enhanced abundance of small grains leads to more efficient
absorption of dissociating radiation.
In contrast, the CO abundance is lower for $\eta_\mathrm{dense}=0.9$ than for $\eta_\mathrm{dense}=0.5$
because the grain sizes biased to larger radii lead to less absorption of dissociating radiation.
Accordingly, $X_\mathrm{CO}$ is larger/smaller for larger/smaller $\eta_\mathrm{dense}$.
Therefore, the CO abundance and CO-to-H$_2$ conversion factor are significantly
affected by the evolution of grain size distribution.

The star formation time $\tau_\mathrm{SF}$, which regulates the time-scale of metal
enrichment, also strongly affects the metallicity dependence of the molecular abundances and
the CO-to-H$_2$ conversion factor. Since dust growth by accretion becomes efficient at
higher metallicity for quicker star formation,
the increase of $f_\mathrm{CO}$ occurs
at higher metallicity for shorter $\tau_\mathrm{SF}$.
As a consequence, the drop of $X_\mathrm{CO}$ occurs at higher metallicity for
shorter $\tau_\mathrm{SF}$.
For $\tau_\mathrm{SF}=0.5$ Gyr, galaxies are CO- and dust-poor up to the point where the
metallicity reaches nearly solar.
This emphasizes the importance of tracing star-forming clouds with [C \textsc{ii}] or
[C \textsc{i}] emission for high-redshift star-forming galaxies, especially at the epoch when the cosmic
age is $\lesssim 0.5$ Gyr.

We also show that, if we plot $X_\mathrm{CO}$ as a function of dust-to-gas ratio instead of metallicity,
all models above for various $\eta_\mathrm{dense}$ and $\tau_\mathrm{SF}$ are aligned along a single
relation. Thus, we propose dust-based formulae for $X_\mathrm{CO}$ including one with
correction for the grain size distribution (equations \ref{eq:Xco_dg} and \ref{eq:Xco_dg_corr}).
However, we should note that an observational estimate of dust-to-gas ratio requires
$X_\mathrm{CO}$. Thus, the dust-based formulae can be used only in an iterative manner, and
their usefulness needs to be checked in future studies.

Our results show that the CO-to-H$_2$ conversion factor is strongly linked to the evolution of
grain size distribution. Thus, our model provides a guide for how to choose $X_\mathrm{CO}$
for the population of galaxies whose evolutionary status of dust is very different from the MW.
We also emphasize that $X_\mathrm{CO}$ is affected by the grain size distribution
even if the metallicity or the dust-to-gas ratio is the same.

\section*{Acknowledgements}

{We thank the anonymous referee for useful comments that improved and deepened the
discussions in this paper.}
We are grateful to I-Da Chiang for useful discussions on the CO-to-H$_2$ conversion factor.
We thank the National Science and Technology Council for support through grants
108-2112-M-001-007-MY3 and 111-2112-M-001-038-MY3,
and the Academia Sinica for Investigator Award AS-IA-109-M02.

\section*{Data availability}
The data underlying this article will be shared on reasonable request to the corresponding author.



\bibliographystyle{mnras}
\bibliography{/Users/hirashita/bibdata/hirashita}

%%%%%%%%%%%%%%%%% APPENDICES %%%%%%%%%%%%%%%%%%%%%

\appendix

\section{Effects of shielding on the CO abundance}\label{app:CO}

\begin{figure}
 \includegraphics[width=\columnwidth]{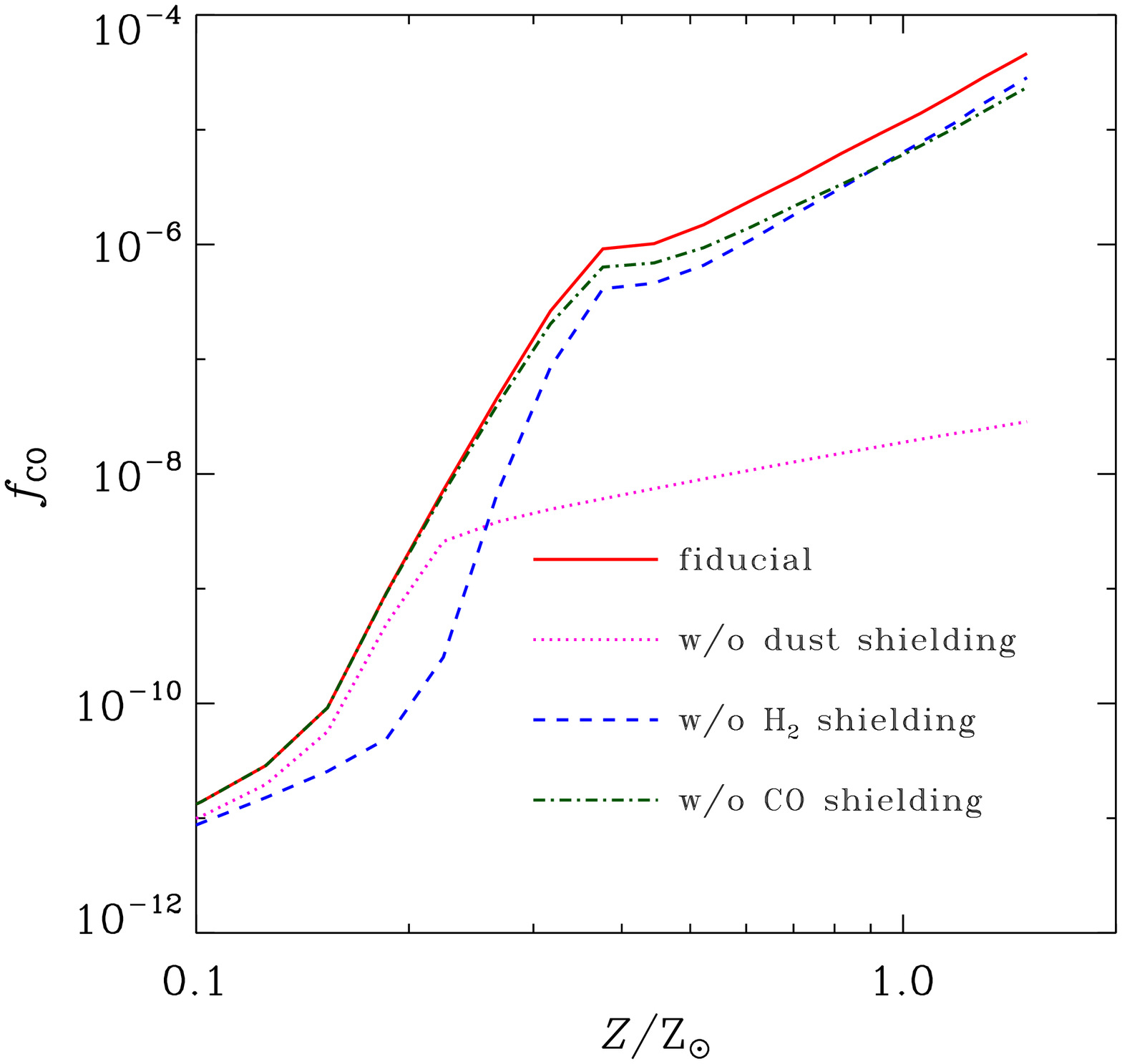}
 \caption{Evolution of the CO fraction, $f_\mathrm{CO}$, as a function
 of metallicity in a typical cloud with the fiducial parameter set.
 The solid line shows the resulting relation with all the
 shielding taken into account, and the dotted, dashed, and dot--dashed
 lines present the relations without shielding by dust, H$_2$, and CO,
 respectively.}
 \label{fig:CO_appendix}
\end{figure}

We show which shielding mechanism is important for the CO abundance $f_\mathrm{CO}$.
To this goal, we calculate $f_\mathrm{CO}$ without one of the shielding sources:
dust, H$_2$, and CO, and the result is compared with the case with all these shielding
mechanisms. We adopt the fiducial parameter values.
The CO abundance is shown as a function of metallicity in Fig.\ \ref{fig:CO_appendix}.

We observe in Fig.\ \ref{fig:CO_appendix} that dust shielding has the largest impact
on $f_\mathrm{CO}$, especially at high metallicity.
Therefore, correctly modelling the dust properties, including the grain size distribution,
is essential in estimating $f_\mathrm{CO}$ and $X_\mathrm{CO}$.
The second most important shielding is caused by H$_2$, which contributes to
a factor 2 increase of $f_\mathrm{CO}$.
The effect of H$_2$ shielding is almost constant at $Z\gtrsim 0.2$ Z$_{\sun}$
because the cloud is already fully molecular. CO self-shielding is not negligible at solar
metallicity, affecting the CO abundance as much as H$_2$ shielding.

% Don't change these lines
\bsp	% typesetting comment
\label{lastpage}
\end{document}